\documentclass[journal,9pt]{IEEEtran}
\usepackage[cmex10]{amsmath}
\usepackage{graphicx}
\usepackage{caption}
\usepackage{subfig}
\usepackage{multicol}
\usepackage{soul}

\usepackage{lipsum}
\usepackage{adjustbox}
\usepackage[figuresright]{rotating}

\usepackage{algorithm}
\usepackage[noend]{algpseudocode}
\usepackage{tablefootnote}

\usepackage{pifont}

\usepackage{multirow}
\usepackage{url}
\usepackage[table,xcdraw]{xcolor}
\usepackage{amssymb}

\usepackage{rotating}
\usepackage{tikz}

\newcommand{\cmark}{\ding{51}}%

\soulregister\cite7
\soulregister\ref7
\soulregister\pageref7

\ifCLASSINFOpdf
   \graphicspath{charts/}
   \DeclareGraphicsExtensions{.pdf,.jpeg,.png}
\else
\fi

\newcommand{\ballnumber}[1]{\tikz[baseline=(myanchor.base)] \node[circle,fill=.,inner sep=1pt] (myanchor) {\color{-.}\bfseries\footnotesize #1};}

\begin{document}
%
\title{{A Modeling Framework for Reliability of Erasure Codes in SSD Arrays}}


\author{\IEEEauthorblockN{Mostafa Kishani, Saba Ahmadian, and Hossein Asadi, Senior Member, IEEE\\}
\IEEEauthorblockA{Data Storage, Networks, \& Processing (DSN) Lab, Department of Computer Engineering \\Sharif University of Technology\vspace{-1.8em}}}


%


\maketitle

\begin{abstract}

Emergence of \emph{Solid-State Drives} (SSDs) have evolved the data storage industry where they are rapidly replacing \emph{Hard Disk Drives} (HDDs) due to their superiority in performance and power. 
{Meanwhile,} SSDs have reliability issues due to bit errors, bad blocks, and bad chips. 
To help reliability, \emph{Redundant Array of Independent Disks} (RAID) 
configurations, originally proposed to increase both
performance and reliability of HDDs, are also applied to SSD arrays. 
However, the conventional reliability models of HDD RAID cannot {be intactly applied to SSD arrays}, 
as the nature of failures in SSDs are totally different from HDDs. 
Previous studies on the reliability of SSD arrays are based on the deprecated SSD failure data, and \emph{only} focus on limited failure types, device failures, and page failures caused by the bit errors, while recent field studies have reported other failure types including bad blocks and bad chips, and a high correlation between failures.

In this {paper}, we investigate the reliability of SSD arrays using field storage traces and real-system implementation of conventional and emerging erasure codes. 
The reliability is evaluated by statistical fault injection experiments that {post-process} the usage logs obtained from the real-system implementation, 
while the fault/failure attributes are obtained from the state-of-the-art field data by previous works. 
As a case study, we examine conventional RAID5 and RAID6 and emerging \emph{Partial-MDS} (PMDS) codes, \emph{Sector-Disk} (SD) codes, and \emph{STAIR} codes in terms of both reliability and performance using an open-source software RAID controller, MD (in Linux kernel version 3.10.0-327), and arrays of Samsung 850 Pro SSDs. 

Our detailed analysis on the data loss breakdown shows that a) emerging erasure codes fail to replace RAID6 in terms of reliability, b) row-wise erasure codes are the most efficient choices for contemporary SSD devices, {and} c) previous models overestimate the SSD array reliability by up to six orders of magnitude, as they just focus on the coincidence of bad pages (bit errors) and bad chips within a data stripe that holds the minority of root cause of data loss in SSD arrays. 
Our experiments show that the combination of bad chips with bad blocks is recognized as the major source of data loss in RAID5 and emerging codes (contributing more than 54\% and 90\% of data loss in RAID5 and emerging codes, respectively), while RAID6 remains robust under these failure combinations.
Finally, the fault injection results {reveal} that SSD array reliability, as well as the failure breakdown is significantly correlated with SSD type.
\end{abstract}


%
\IEEEpeerreviewmaketitle

\vspace{-0.3cm}
\section{Introduction}
\begin{figure}
    \centering
        \includegraphics[width=0.45\textwidth]{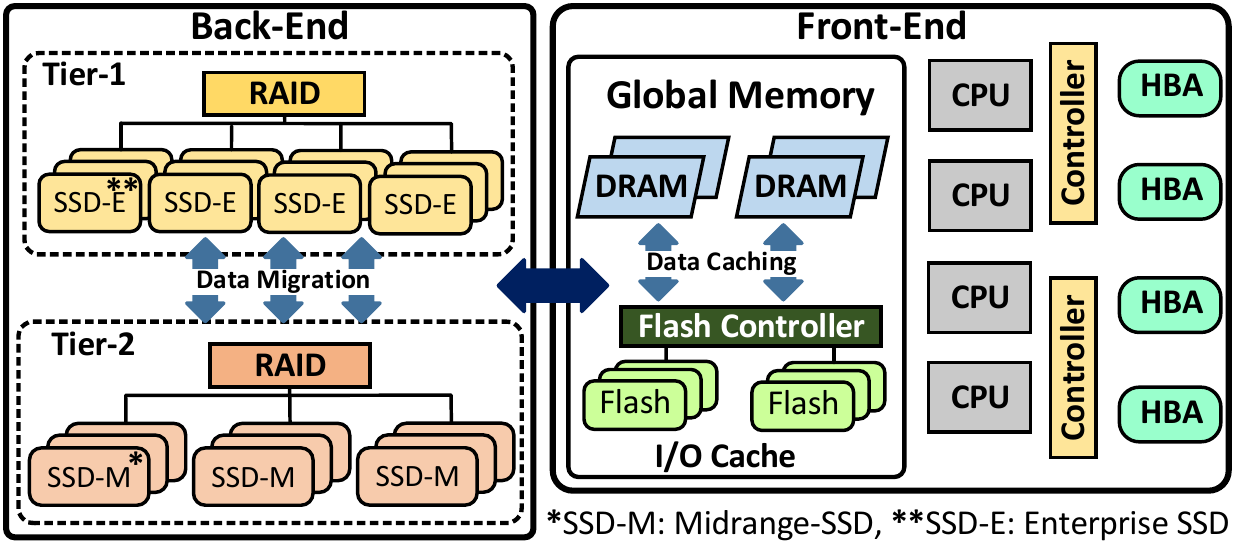}
    \caption{Structure of an enterprise {flash-based} storage system.}
    \label{fig:all-flash-structure}
 \vspace{-0.2cm}
\end{figure}

\emph{Solid-State Drives} (SSDs) are predicted to replace \emph{Hard Disk Drives} (HDDs) due to their performance and power consumption benefits~\cite{demara2018non}. 
While a big spectrum of \emph{Non-Volatile Memory} (NVM) technologies are appeared in the recent years and struggle to find their place in industry~\cite{wang2018rc,henkel2017emerging,huai2008spin,yang2014improving,yang2011flexible,khoshavi2018read,fedorov2017speculative,qin2011mnftl,liu2011pcm,qin2010demand,qin2011two,dong2014nvsim,wu2009hybrid,joo2010energy,zand2017energy,guan2017block,kang2018reliability,wang2015propram,zhao2016optical,salkhordeh2019analytical}, SSDs are still the most matured and promising high-performance storage devices. 
{SSDs are intensively used a) as the main} storage media in all-flash {storage systems}, {b)} as a caching media in \emph{Input/Output} (I/O) cache layer
~\cite{li2014nitro,li2017pannier,liu2013duracache,li2014nitro,tarihi2016hybrid,reca,ahmadian2018eci,salkhordeh2018efficient}, 
and {c)} for {tiering} purposes~\cite{hu2011performance,guerra2011cost,salkhordeh2015operating} (Fig.~\ref{fig:all-flash-structure}). 
Meanwhile, SSDs have reliability issues due to wear-out\footnote{Each flash cell can tolerate a limited number of writes (erasures) and wears out after a few thousand of erasures, depending on the technology and device variations.}, bit errors, bad blocks, and bad chips. These reliability issues can increase the chance of data unavailability and data loss in storage systems~\cite{elerath2014beyond,park2009reliability,kishani2019dependability,kishani-tr-2018}. 

To enhance reliability, \emph{Redundant Array of Independent Disks} (RAID) 
configurations~\cite{patterson1988case} are employed in storage systems to avoid data loss and data unavailability. 
However, the nature of failures and errors in SSDs are totally different from
HDDs~\cite{schroeder2016flash,meza2015large,narayanan2016ssd,ahmadian-ssd-rel-date}; SSDs have increasing \emph{Bit Error Rate} (BER) with a distribution different
from HDDs and there is a high correlation between bit errors in SSDs~\cite{schroeder2016flash,meza2015large,narayanan2016ssd}.
Due to these differences, conventional reliability models of HDD arrays cannot be {applied intactly to SSDs}. Previous studies on the reliability of SSD arrays~\cite{li2016analysis,kim2013improving,moon2016does,balakrishnan2010differential,greenan2009building,blaum2013partial}, however, are based on old SSD failure studies~\cite{grupp2012bleak,cai2012error,grupp2009characterizing}, and just focus on a single failure type, page failures caused by bit errors, while recent field studies have reported other failure types including bad blocks and bad chips alongside page failures, and a high correlation between these failure types.

A recent work by Schreoder et al.~\cite{schroeder2016flash} reports SSD failure data in Google datacenters and shows that alongside bit errors that result in the loss of one page of data, device failures, including bad blocks and bad chips also affect the reliability of SSD arrays, despite previous work that just reports \emph{Raw Bit Error Rate} (RBER)\footnote{RBER is defined as the number of corrupted bits over the total number of read bits (including both correctable and uncorrectable errors)~\cite{schroeder2016flash}.} and \emph{Uncorrectable Bit Error Rate} (UBER). 
This study also reports a high correlation between RBER and parameters such as prior \emph{Program/Erase} (P/E) cycles, and a high correlation between bad chip and total number of bad blocks in a SSD chip. 
Meza et al.~\cite{meza2015large} and Narayanan et al.~\cite{narayanan2016ssd} have also reported SSD failure field data in the recent years.
The reports of all the mentioned works contradict the old SSD failure studies~\cite{grupp2012bleak,cai2012error,grupp2009characterizing}, hence, discredit existing reliability models that are based on those data. 
Many studies try to model the reliability of SSD arrays and modify RAID configurations in favor of SSD failure characteristics
~\cite{li2016analysis,kim2013improving,moon2016does,balakrishnan2010differential,plank2014sector,li2014stair}. These works, however, come with very trivial or no reliability estimations, \emph{{or}} are based on misleading SSD failure characteristics reported by old studies~\cite{grupp2012bleak,cai2012error,grupp2009characterizing}.
To the best of our knowledge, \emph{none} of previous studies have modeled SSD array reliability based on valid field data and real-system implementation. 

In this {paper}, we investigate the reliability of SSD arrays using the real-system implementation of conventional RAID and emerging erasure codes. 
The reliability is evaluated by statistical fault injection experiments that {post-process} the SSD usage logs obtained from the system run, 
while the fault/failure attributes are obtained from the state-of-the-art field data by previous works~\cite{schroeder2016flash}. 
As a case study, we examine conventional RAID5 and RAID6 and emerging \emph{Partial-MDS} (PMDS) codes~\cite{blaum2013partial}, \emph{Sector-Disk} (SD) codes~\cite{plank2014sector}, and \emph{STAIR} codes~\cite{li2014stair} in terms of reliability, endurance, and performance.
The experiments are conducted using an open-source software-RAID controller, \emph{Multiple Device} (MD), in Linux kernel version 3.10.0-327 (CentOS 7 operating system), and arrays of Samsung 850 Pro SSDs.

{Thorough investigation of SSD arrays has revealed the following major observations:
	1) Erasure codes mainly suffer from data loss caused by the combination of device failure and block failure. In the RAID5 arrays, bad chips combined with either bad pages or bad blocks are the major sources of data loss. 
	The contribution of bad blocks combined by bad chip (two bad blocks and one bad chip) is also significant in total data loss of RAID6.
	2) Unlike previous models which only focus on the coincidence of bad pages and bad chips, our study shows that this type of failure contributes the minority of data loss in SSD arrays.
	3) SSD array reliability, as well as failure breakdown is significantly correlated with SSD type.
	4) Time to scrub has a significant impact on array reliability, while the impact of time to recover from a device failure is of less significance. 
	5) RAID5 and RAID6 codes perform almost independent of stripe size. Emerging erasure codes, however, benefit from smaller stripe sizes and show a promising reliability improvement when reducing stripe size.}

We can summarize the major contributions of this work as follows:
\begin{itemize}
\item
We propose an analytic model for the reliability of SSD arrays, considering realistic SSD failure attributes from state-of-the-art studies in the field.
\item
We propose a generalized fault injection framework for evaluating the reliability of SSD arrays, using SSD usage logs obtained by real-system implementation and SSD failure characteristics obtained by state-of-the-art field data.
\item
We evaluate the reliability of different erasure codes under an extensive number of representative storage workloads. 
\item
We compare the performance and endurance overhead of different erasure codes using the real storage stack, despite previous works that inadequately compare decode/encode complexity and ignore the endurance and {I/O} overhead. 
\item
We {develop} an open-source framework for SSD array fault injection, 
{which} will be publicly available for the research {community.}\footnote{The framework is available in \url{http://dsn.ce.sharif.edu/}} 
\end{itemize}

The rest of {this} paper is organized as follows. 
Section~\ref{sec:related} discusses related work on SSD reliability.
Section~\ref{sec:background} presents a background about examined erasure codes.
Section~\ref{sec:proposedmodel} discusses the proposed modeling framework. 
Section~\ref{sec:Results and Observations} presents the experimental setup, results, and the corresponding observations and discussions.
Finally, Section~\ref{sec:conclusion} concludes the paper.

\section{Related Work}
\label{sec:related}

\subsection{Field Studies on SSD Failure Characteristics}
A recent work by Schreoder et al.~\cite{schroeder2016flash} investigates SSD reliability by collecting six-year SSD failure data in Google datacenters. 
This study shows that alongside bit errors that result in the loss of one page of data, device failures, including bad blocks and bad chips are also of major reliability threats in SSD arrays, despite previous work that just report RBER and UBER. 
This study also reports a high correlation between RBER and parameters such as prior P/E cycles, SSD age, read count, write count, erase count, and prior RBER. 
Another field study by Meza et al.~\cite{meza2015large} reports that RBER does not monotonically increase with P/E cycles and also reports that RBER has an exponential growth in SSD useful life. However, this study reports a smooth linear increase of RBER with P/E cycles. 
This study also shows a high correlation between total number of bad blocks in a SSD chip and the number of bad blocks already developed. It also shows that in an over four year mission, more than 30-80\% of SSDs experience bad blocks in the field.
Another observation of this study is that 2-7\% of SSDs experience bad chip within the first four year of their life. 

A work by Meza et al.~\cite{meza2015large} conducts a deep study on the failure characteristics of flash memories using field data from Facebook datacenters.
This work observes that SSD failure rate does not increase monotonically with flash chip wear. 
In turn, SSD failure rate has four phases of early detection, early failure, useful life, and wear-out~\cite{meza2015large}. 
Another observation is that UBER obtained in this work is 10 to 1000 times smaller than the raw BER of similar flash chips examined by Grupp et al. ~\cite{grupp2012bleak}. This is due to the fact that SSDs in this work correct small errors, perform wear leveling, and are not at the end of their rated life~\cite{meza2015large}.
Meza et al. {show} that on average 10\% of SSDs experience 80\% of all uncorrectable errors, while in most of platforms 10\% of SSDs experience 95\% of all observed uncorrectable errors. 
It also shows that during two successive weeks, 98\% of SSDs that experienced an error during the first week also had an error during the next week.   

Grupp et al.~\cite{grupp2012bleak} also report BER for \emph{Single Level Cells} (SLC) and \emph{Multi Level Cells} (MLC) flash of different feature size. 
In another research, Grupp et al.~\cite{grupp2009characterizing} {show} that BER increases by flash chip wear. However, this work does not consider the effect of optimizations in the SSD controller and buffering layers. 
Cai et al.~\cite{cai2012error} also examine the bit error patterns in MLC NAND flash and demonstrate its dependency to P/E cycles, physical location, and value.  
Finally, Mielke et al.~\cite{mielke2008bit} report BER and sector failure of MLC NAND flash in conjunction with P/E cycles, retention time, and number of reads. 

\subsection{Analysis and Modeling of SSD Array Reliability}
A large body of research has investigated and tried {to improve} the reliability of disk arrays~\cite{greenan2009building,greenan2007disaster,paris2007self,rincon2017disk,schwarz2016resar,paris2013three,kao2013flexible,gopinath2009reliability,paris2008mttdls,kishani-tr-2018,kishani2017}. For the sake of brevity, here we focus on the studies concentrating on \emph{SSD} arrays. 
Greenan et al.~\cite{greenan2009building} propose a combination of inter-device and intra-device parity codes to cope with page failures, block failures, and device failures in SSD arrays. 
While the authors have a realistic assumption about failure types in SSD arrays, their reliability assessment approach is questionable, as it reports \emph{Uncorrectable Page Error Rate} (UPER) using cumulative Binomial distribution as a function of RBER.
The proposed method also necessitates the migration of \emph{Flash Translation Layer} (FTL) from device to RAID controller. Hence, the method cannot be employed using \emph{Commercial off-the-Shelf} (COTS) devices.
 
Balakrishnan et al.~\cite{balakrishnan2010differential} propose Differential RAID as an alternative to conventional RAID5 to be applied in SSD arrays. The idea is based upon uneven parity distribution across array devices (in the most intense configuration, RAID5 is transformed to RAID4) to reduce time proximity of wear-out phenomenon in SSDs. 
This method is examined using statistical fault injections.
Li et al.~\cite{li2016analysis} compare RAID5 with Differential RAID~\cite{balakrishnan2010differential} in SSD arrays. 
The work has a mathematical discussion, adopted from~\cite{e2000transient}, to apply a variable failure rate to \emph{Continuous Time Markov Chain } (CTMC) using \emph{Kolmogorov} forward equation, \emph{uniformization}~\cite{e2000transient}, and \emph{truncation}.
This work validates the mathematical model by statistical fault injections using Microsoft SSD simulator~\cite{agrawal2008design} extended from DiskSim~\cite{bucy2008disksim}, and estimates the reliability as a function of erasures (SSD age).
One important shortcoming of this work is considering equal failure rate for all devices using Weibull distribution of SSD bit error rate, and ignoring the correlation of errors.

Kim et al.~\cite{kim4ds} {attempt} to improve the reliability of RAID5 in SSD arrays by proposing 
\emph{Dynamic Striping-RAID} (DS-RAID). 
This work compares the proposed method with conventional RAID5 in terms of response time and number of write operations, including original data writes and extra writes due to parity and garbage collection, as a representative for SSD lifetime. 
Finally, Kim et al.~\cite{kim2013improving} propose \emph{Elastic Striping and Anywhere Parity} (eSAP-RAID) as an alternative to RAID5 with {higher} performance and reliability in SSD arrays. 
This method tries to reduce the number of writes due to parity updates, by allowing flexible stripe size and parity placement. Both works~\cite{kim4ds,kim2013improving} use Microsoft SSD simulator~\cite{agrawal2008design} extended from DiskSim~\cite{bucy2008disksim}.  

Moon and Reddy~\cite{moon2016does} investigate the reliability of RAID0, RAID1, and RAID5 in SSD arrays by considering the effect of garbage collection and show the trade-off between reliability and utilization in a SSD array. 
This work arguably uses Markov models by considering constant bit error rate for SSDs and ignores the correlation between errors.
Blaum et al.~\cite{blaum2013partial} {propose} a new family of erasure codes to cope with the coincidence device failures and symbol (page) failures.
The proposed code is evaluated using probabilistic analysis of data loss.
In summary, Table~\ref{tab:qualitative-comparison} makes a qualitative comparison between different SSD array reliability models and our proposed framework. 
As the table shows, unlike our proposed framework, previous works do not use real-system implementation and also make use of deprecated SSD failure data.

\begin{table}
\centering
\caption{Qualitative comparison of proposed framework with different SSD array reliability models by Balakrishnan et al.~\cite{balakrishnan2010differential}, Blaum et al.~\cite{blaum2013partial}, Li et al.~\cite{li2016analysis}, and Moon and Reddy~\cite{moon2016does}.}
\vspace{-0.3cm}
\begin{center}
\begin{adjustbox}{width=0.5\textwidth,totalheight=\textheight,keepaspectratio}
    \begin{tabular}{ | l | c | c | c | c | c |}
    \hline
&	\footnotesize{~\cite{balakrishnan2010differential}}	&	\footnotesize{~\cite{blaum2013partial}}&	\footnotesize{~\cite{li2016analysis}}&	\footnotesize{~\cite{moon2016does}}&	\footnotesize{Proposed} \\ \hline \hline
Coincidence of device and symbol failure &	\cmark	& \cmark & 	& \cmark	&\cmark \\ \hline
SSD age	& \cmark	& 	& \cmark	& 	& \cmark  \\ \hline
BER as a function of erasures	& \cmark	&	&\cmark	&	&\cmark \\ \hline
Statistical fault injection & \cmark & & \cmark &  & \cmark \\ \hline
Use mathematical reliability analysis & & \cmark & \cmark & \cmark &  \\ \hline
Emerging erasure codes & & \cmark & & & \cmark \\ \hline
Valid SSD failure field data	&	&	&	&	&\cmark \\ \hline
Correlation between errors	&	&	&	&	&\cmark \\ \hline
Valid distribution for BER	&	&	&	&	&\cmark \\ \hline
Real-system implementation 	&	&	&	&	&\cmark \\ \hline
    \end{tabular}
    \end{adjustbox}
\end{center}
\vspace{-0.6cm}
\label{tab:qualitative-comparison}
\end{table}

\section{Background}
\label{sec:background}

\begin{figure}[!htb]
	\centering
	\subfloat[Linear parity calculation]{\includegraphics[width=.24\textwidth]{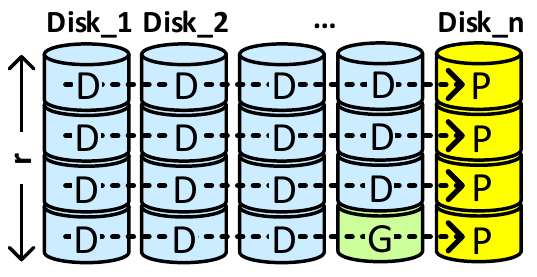}%
		\label{fig:pmds-linear-parity}}
	\hfil
	\hspace{-0.4em}
	\subfloat[Global parity calculation)]{\includegraphics[width=.24\textwidth]{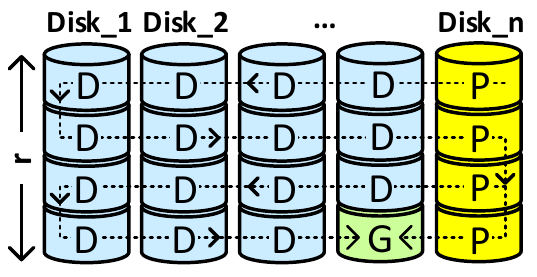}%
		\label{fig:pmds-global-parity}}
	
	\caption{Structure of a data stripe using $PMDS(1,1)$ code when the number of devices and rows are respectively $n$ and $r$. {Note} $D$, $P$, and $G$ respectively stand for data symbol, parity symbol, and Global parity symbol.}
	\label{fig:pmds-stripe-structure}
\end{figure}

RAID is proposed as a solution to cope with performance 
and reliability issues of single disks~\cite{patterson1988case}. 
RAID5 and RAID6 configurations distribute the data to an array of disks
while keeping the row-wise parity of devices in respectively one and two redundant devices. Hence, RAID5  and RAID6 can respectively tolerate one and two device failures. 
We can put both RAID5 and RAID6 in the category of \emph{Maximum Distance Separable} (MDS) codes that offer the maximum correction capability, due to having the maximum hamming distance\footnote{MDS codes have a big spectrum of alternatives such as Reed-Solomon~\cite{wicker1995error}, multi-dimensional codes~\cite{kishani2011hvd}, and simple linear hamming codes usually used in fast memory structures~\cite{kishani2011using}.}. 
Blaum et al.~\cite{blaum2013partial} {propose} PMDS codes to handle the combination of both device failures and symbol (page) failures, by using the combination of row-wise parity and a new concept of \emph{Global Parity} (GP) that is taken across the whole data stripe (Fig.~\ref{fig:pmds-stripe-structure}).  

\vspace{-0.2cm}

Fault tolerance of PMDS codes can be specified by $m$, number of tolerable drive failures (i.e., number of coding drives) and $s$, number of tolerable sector failures (equal to the number of {global parities}). 
For example, $m=1$ and $s=1$ says that one drive failure is tolerable while one of operating chunks can tolerate one sector failure.  
A specific configuration of PMDS codes is capable of tolerating one device and one sector failure, denoted as $PMDS(1,1)$, that is examined in our study.
SD codes~\cite{plank2014sector} and STAIR {codes}~\cite{li2014stair} also propose different methods for encoding and decoding of {global} parity with different computational complexities (but the same I/O overhead)~\cite{li2014stair}.

We assume that each data stripe is composed of $n$ devices (or $n$ data chunks), including redundant devices, and $r$ rows, where $r$ stands for the number of symbols from each device in one stripe.  
Fig.~\ref{fig:pmds-stripe-structure} shows the structure of PMDS(1,1) while the data symbols are denoted by $D$, row-vise parity symbols are denoted by $P$, and the {global} parity symbol is denoted by $G$.
PMDS codes, SD codes and STAIR codes are systematic (separable) codes with homomorphic property. 
This property enables updating the codeword when the data is partially updated by an approach similar to updating normal parity bits, as shown in Equation~\ref{equ:symbol-update}.
\begin{equation}
\label{equ:symbol-update}
\begin{split}
Codeword_{new} = Codeword_{old} \oplus Data_{old} \oplus Data_{new} 
\end{split}
\end{equation} 

Hence, PMDS(1,1), SD(1,1){, and} STAIR(1,1) codes perform similar in terms of I/O penalty, but are different in encoding/decoding computational complexity~\cite{li2014stair}. 
However, as encoding/decoding computation time is negligible compared to I/O penalty, in the rest of this work we note PMDS(1,1), SD(1,1), and STAIR(1,1) codes by PMDS(1,1) or simply PMDS.
Table~\ref{tab:erf} shows the \emph{Effective Replication Factor} (ERF) and computations (XORs) needed for encoding one stripe of RAID5, RAID6, and PMDS (discussed in detail by Li and Lee~\cite{li2014stair}). 
We also compare the I/O penalty of erasure codes in Table~\ref{tab:update penalty comparison}.
{This table} shows that in the case of sector and row update, PMDS has more number of write/read compared to RAID5 and RAID6, while in the case of stripe update both RAID5 and PMDS have an equal overhead lower than the overhead of RAID6. 
{This analysis shows that in sequential workloads dominated by full stripe writes, we can expect a greater performance from PMDS, compared to RAID6. 
We further verify this hypothesis by examining different realistic workloads.}

\begin{table}
\centering
\caption{ERF and {computation} (number of XORs) needed for updating one data stripe of different erasure codes.}
\vspace{-0.3cm}
\begin{center}
    \begin{tabular}{ |c| c | c | c |}
    \hline
    &\textbf{RAID5} & \textbf{RAID6} & \textbf{STAIR(1,1)}  \\ \hline \hline
	XORs& $(n-1) \times r$ & $(n-1) \times r \times 2$ & $(n-1) \times r \times 2 + r-1$ \\ \hline
	ERF&$\frac{n+1}{n} $ & $\frac{n+2}{n}$ & $\frac{(n+1)\times r}{(n \times r) - 1}$   \\ \hline
    \end{tabular}
\end{center}
\label{tab:erf}
\end{table}

\begin{table}
\centering
\caption{Write/read operations needed for updating one page, one row, and one data stripe. $W$ stands for the write operation, and $R$ stands for read before write.}
\vspace{-0.3cm}
\begin{center}
\begin{tabular}{|c|c|c|c|}
\hline
                               & RAID5 & RAID6 & PMDS(1,1) \\ \hline
\multirow{2}{*}{Sector Update} &    W=2   &   W=3    &   W=4         \\ \cline{2-4} 
                               &    R=2   &    R=3   &     R=4       \\ \hline
\multirow{2}{*}{Row Update}    &  W=$n+1$     &    W=$n+2$   &    W=$n+3$        \\ \cline{2-4} 
                               &   R=0    &    R=0   &       R=$n+2$     \\ \hline
\multirow{2}{*}{Stripe Update} &  W=$(n+1) \times r$     & W=$(n+2) \times r$      &  W=$(n+1) \times r$          \\ \cline{2-4} 
                               &  R=0     & R=0       & R=0           \\ \hline
\end{tabular}
\end{center}
\vspace{-0.5cm}
\label{tab:update penalty comparison}
\end{table}

\section{Modeling Reliability of RAID5, RAID6, and PMDS Codes in SSD arrays}
\label{sec:proposedmodel}

We model the reliability of SSD array for different erasure codes, by proposing a fault injection environment that uses the field data of SSD failure statistics from Schreoder et al.~\cite{schroeder2016flash}, alongside SSD operation log from 
arrays of Samsung 850 Pro SSDs, using the open-source software RAID controller MD in Linux kernel version 3.10.0-327.
We consider three possible failure types in SSD arrays, reported by field studies~\cite{schroeder2016flash}, including \emph{Bad Page}, \emph{Bad Block}, and \emph{Bad Chip}. 
\begin{itemize}
\item
\emph{\textbf{Bad Page}} \textbf{($BP$)} or \emph{\textbf{Bad Symbol}} \textbf{($BS$)} is the most prevalent failure type in an SSD array, caused by uncorrectable bit errors in SSD device. As in the storage systems, data is logically read/written/managed in the units of pages, the page is considered lost when it contains uncorrectable corrupted bits. 
We call this failure \emph{Bad Page} or \emph{Bad Symbol}, as the page is the smallest data symbol that different erasure codes are performed on.  
A bad symbol can result in the loss of one data stripe, if it is not correctable by the  employed erasure code.
Note bit errors that are correctable by the internal \emph{Error Correction Code} (ECC) of the SSD device are not considered as bad symbol. 
\item
\emph{\textbf{Bad Block}} \textbf{($BB$)} is reported by field studies as another common failure type in SSD arrays~\cite{schroeder2016flash}. 
As each block contains multiple (tens or hundreds) pages, uncorrectable bad blocks can affect multiple data stripes in the SSD array, depending on the 
data striping protocol.  
\item
\emph{\textbf{Bad Chip}} \textbf{($BC$)} is the last type of failure in the SSD arrays, reported by field studies~\cite{schroeder2016flash}. 
Bad chip can result in the loss of whole array, for example when two bad chips on two different array devices coincide in the case of RAID5. 
It also can result in the loss of one or multiple data stripes, when it coincides with a bad symbol or bad block in another array device in the case of RAID5. 
\end{itemize}

\vspace{-0.3cm}
\subsection{Correction Capability of RAID5, RAID6, and PMDS}
  
  \begin{figure}
    \centering
        \includegraphics[width=0.5\textwidth]{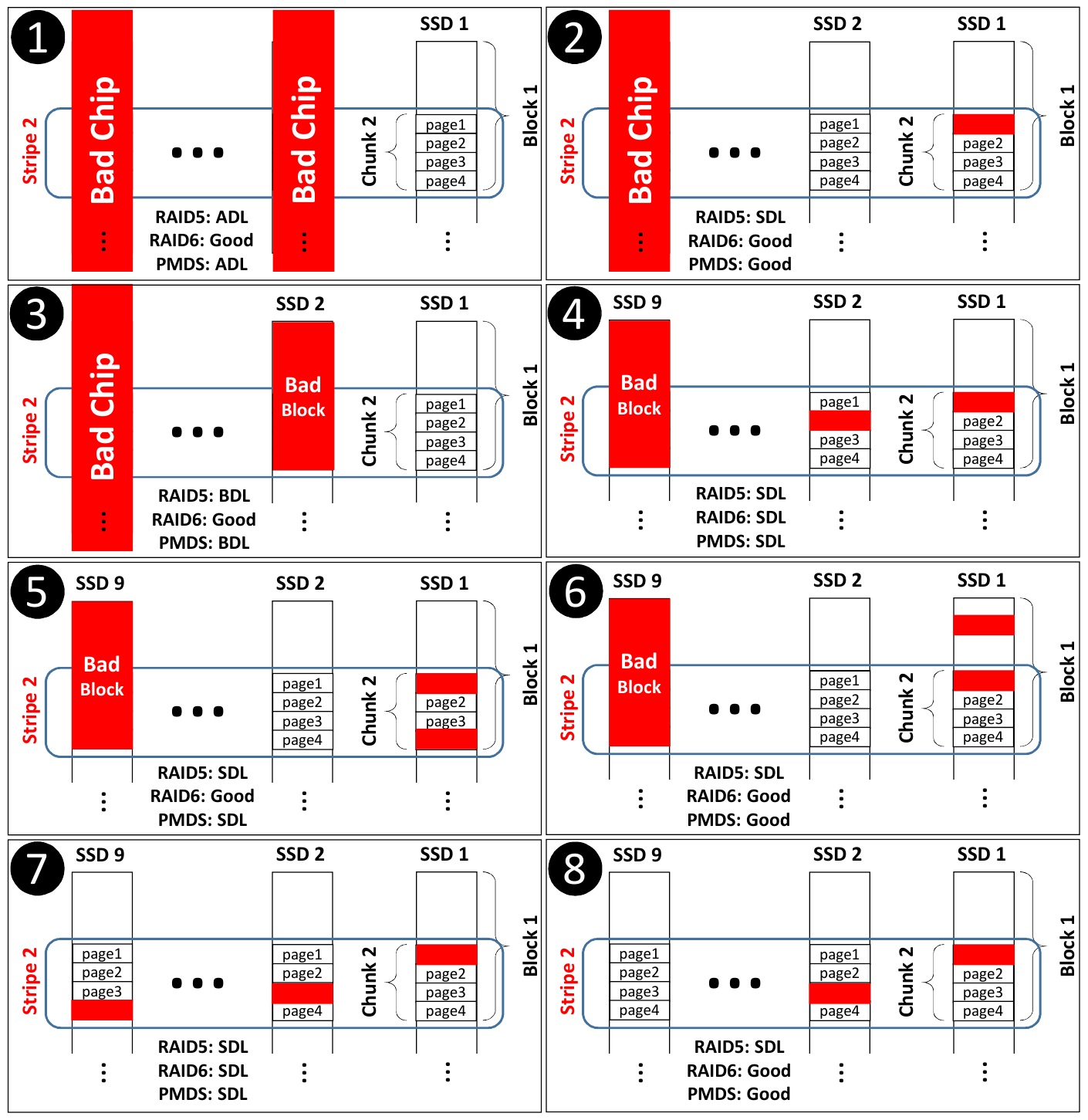}
    \caption{Failure  samples in different SSD arrays.}
    \label{fig:failure-samples}
 \vspace{-0.5cm}
\end{figure}
  
Fig.~\ref{fig:failure-samples} shows eight examples of the combination of bad symbol, bad block, and bad chip in an SSD array.  In this figure, \emph{Array Data Loss} {(ADL)} stands for the loss of whole SSD array, \emph{Block Data Loss} (BDL) stands for the loss of 
all data stripes a block is shared upon, \emph{Stripe Data Loss} (SDL) stands for the loss of one data stripe and \emph{Good} stands for no data loss. 
In this {figure,} we consider a fully striped SSD array, in which a stripe is composed of data chunks from all SSD devices. 
Without loss of generality, here we assume each stripe contains the data from one chip of each SSD device.
Taking other assumptions may affect the magnitude of data loss upon failure incidence.
Each data chunk contains multiple pages (4 in this example) and each block is shared upon multiple stripes (2 in this example).
We should note that erasure codes are performed on the stripe unit, hence, the uncorrectable loss of a single page is considered as the loss of whole stripe (SDL).  

{In example} \ballnumber{1}{, coincidence of two bad chips result in ADL in RAID5 and PMDS while it is recoverable in the case of RAID6. In example} \ballnumber{2}{, the combination of one bad chip and  one bad symbol is correctable in the case of RAID6 and PMDS, but it results in SDL in RAID5. The combination of bad chip and bad block in example }\ballnumber{3} {results in BDL in the case of RAID5 and PMDS, as bad block affects the entire data chunk rather than one symbol. In example} \ballnumber{4}, {all erasure codes face SDL, as three data chunks are corrupted. 
In example} \ballnumber{5}{, RAID5 experiences SDL due to the combination of a bad block and a bad symbol. PMDS also experiences SDL in example }\ballnumber{5}{, as it cannot tolerate more than one symbol failure coincided with a bad chip. Example} \ballnumber{6}{ is similar to example }\ballnumber{5}{, however, in this case PMDS can correct the failure incidence, as two symbol failures }{occur in two different stripes. The coincidence of three bit errors in three different symbols in example }\ballnumber{7}{ also results in SDL in all erasure codes. Finally, in example }\ballnumber{8}{ where two bit errors in different symbols coincide, PMDS and RAID6 can correct the failure, but RAID5 experiences SDL as it cannot tolerate multiple symbol failures in multiples chunks of a single stripe.
Here we can conclude that PMDS can tolerate multiple symbol failures in one data chunk alongside a single symbol failure in another data chunk. 
RAID6 can tolerate multiple symbol failures in two data chunks, and RAID5 can tolerate multiple symbol failures in one data chunk.}

\subsection{Analysis of RAID5, RAID6, and PMDS Reliability}
\label{sec:analysis erasure codes}

Fig.~\ref{fig:reliability-state-diagram} shows the state diagram of SSD array reliability, using different erasure codes, from error-free operation to failure incidence (ADL, BDL, and SDL). This analysis is used in our statistical fault injections to evaluate the reliability of different erasure codes.
Field studies show that the failure characteristics of a SSD {will change} when it wears out, i.e., it passes its P/E Limit or \emph{Wear Out Limit} (WOL)~\cite{schroeder2016flash}. 

{Storage systems may have different regulations when they face a worn-out SSD, such as replacing the SSD or continuing its operation up to the failure. However, in this work we model the most conservative assumption that worn-out SSD is replaced with a brand-new one. 
Replacing the worn-out SSD is also performed with two different regulations: 
1) The first regulation removes the worn-out SSD, immediately replaces it with the brand-new SSD, and reconstructs the data of worn-out SSD on the brand-new one using the parity of the other SSDs. This approach, however, may result in data loss when a bad symbol exists in other operating SSDs. The reason is that the data of those stripes containing bad symbols cannot be reconstructed when the worn-out device is removed (and its data is unavailable) in the case of RAID5.  
2) An alternative approach that prevents this data loss case is adding the brand-new SSD when the worn-out SSD is still operational, making a RAID1 configuration between the brand-new and worn-out SSD, waiting for all data of worn-out SSD to be copied into the brand-new one, and finally removing the worn-out SSD. In this study, for the sake of reliability we take the second approach.}

\subsubsection{RAID5}
Fig.~\ref{fig:raid5-reliability} shows the state diagram of RAID5 SSD array reliability. 
In the $OP$ state, none of SSDs have bad symbol, bad block, or bad chip. 
When a SSD device wears out, the array moves to the $OP_{WO}$ state in which one (or multiple) SSD device is worn out and is waiting to be replaced with brand-new one. 
We dedicate $OP_{WO}$ state from $OP$ state as field studies show that SSD failure characteristics change after wear out~\cite{schroeder2016flash}.
If we neglect this change, $OP$ and $OP_{WO}$ states can be merged (the same happens to other $WO$ states). 

Upon a bad chip in $OP$ and $OP_{WO}$ states, the array moves to $EXP\_BC$ state.
In this state (and also $EXP\_BC_{WO}$ state when the array has worn out SSDs), the array moves back to operational state when the failed device is replaced and reconstructed on the brand-new one. However, any successive bad chip, bad block, and bad symbol results in data loss and moves the array to $ADL$, $BDL$, and $SDL$ states, respectively. 
An operational array (in {either} $OP$ or $OP_{WO}$ states) moves to $EXP\_BB$ state when it faces a bad block. In this state (and also $EXP\_BB_{WO}$ state when the array has worn out SSDs), a successive bad chip (in another device) and coincidence of another bad block in the same stripe, called \emph{Same Stripe Bad Block} (SSBB), results in $BDL$.
Moreover, coincidence of a symbol failure in the same stripe, called \emph{Same Stripe Bad Symbol} (SSBS), results in $SDL$. 
In $EXP\_BB$ and $EXP\_BB_{WO}$ states, the chip containing bad block is prone to fail (bad chip) that moves the array to {either} $EXP\_BC$ {or} $EXP\_BC_{WO}$ states, when no other chip contains bad block or bad symbol. 
In $EXP\_BB$ and $EXP\_BB_{WO}$, bad block is detected by a final read error, {write error}, or erase error~\cite{schroeder2016flash}, and removed by reallocating the corrupted block to a safe block that moves the array back to operational state. 

Finally, an operational array (in $OP$ or $OP_{WO}$ states) moves to $EXP\_BS$ state when it faces a bad symbol. 
In $EXP\_BS$ state (and also $EXP\_BS_{WO}$ state when the array has worn out SSDs), 
a successive bad chip, SSBB, and SSBS results in stripe data loss and moves the array to SDL state. 
However, in $EXP\_BS$ and $EXP\_BS_{WO}$ states, bad symbol can be detected by read error or scrubbing~\cite{schwarz2004disk} and be removed by reconstructing the data from parity of other devices, that moves the array to the operational state. 

\subsubsection{RAID6}
Fig.~\ref{fig:raid6-reliability} shows the state diagram of RAID6 SSD array reliability.
As the figure shows, RAID6 can tolerate an extra failure compared to RAID5, due to having  two redundant devices. 
The description of states and transitions is similar to RAID5. The only difference is six states ($OP\_BC$, $OP\_BB$, $OP\_BS$, $OP\_BC_{WO}$,  $OP\_BB_{WO}$, and $OP\_BS_{WO}$ states) added to RAID6 diagram. These states have the same transitions as $EXP$ states in RAID5 diagram, by this difference that a successive failure in $OP$ {state} moves the array to $EXP$ state rather than data loss state.

\subsubsection{PMDS} 
\begin{figure}
    \centering
 
	\subfloat[RAID5] 
	{
	\vspace{-0.6cm} 
        \includegraphics[width=0.5\textwidth]{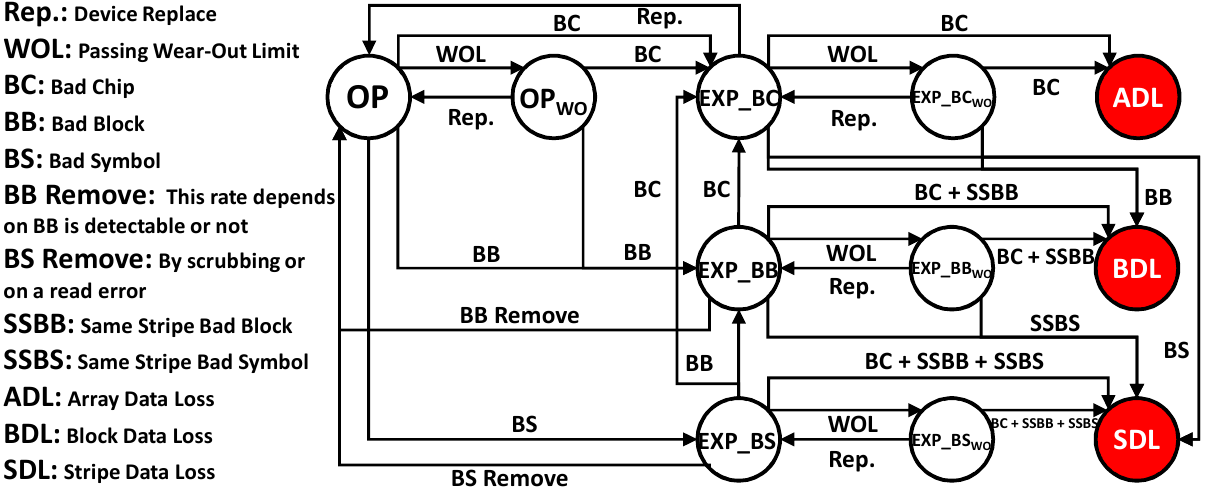}
        \label{fig:raid5-reliability}
    }
    \\
    \subfloat[RAID6]
	{
		\includegraphics[width=0.5\textwidth]{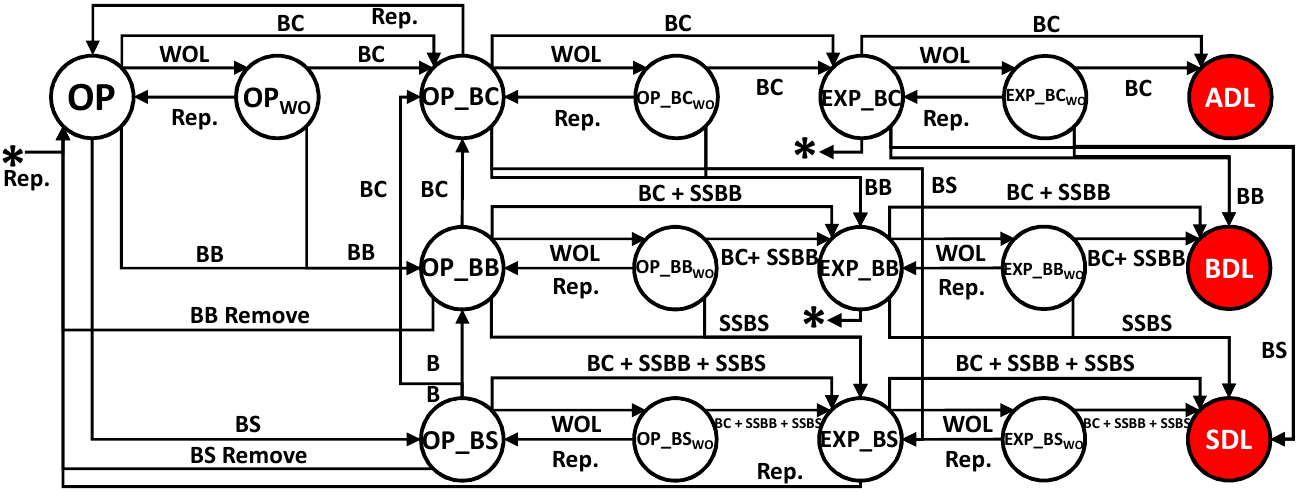}
        \label{fig:raid6-reliability}
    }
    \\
    \subfloat[PMDS Code]
	{     
		\includegraphics[width=0.5\textwidth]{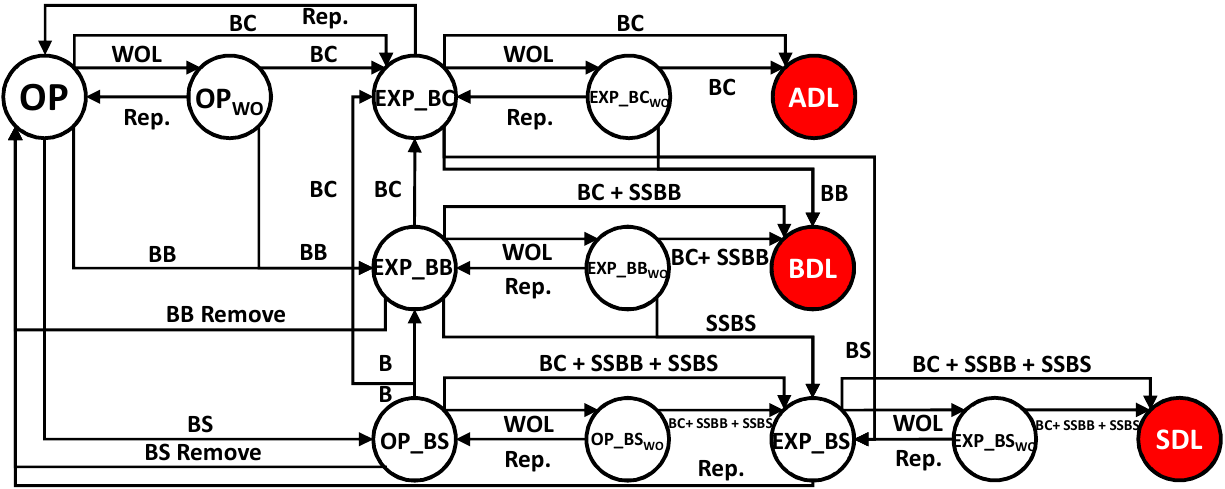}
        \label{fig:sdstair-reliability}
    }
   
    \caption{State diagram of SSD array reliability for RAID5, RAID6, and PMDS code.}
    \label{fig:reliability-state-diagram}
\end{figure}

Fig.~\ref{fig:sdstair-reliability} shows the state diagram of SSD array reliability when employing  PMDS code.
As the figure shows, PMDS performs the same as RAID5 in the case two bad chips coincide (resulting $ADL$) and the case one bad chip coincides with bad block ($BDL$). 
The difference is that PMDS can handle the coincidence of bad chip with bad symbols, and coincidence of two bad symbols in one stripe.
Our fault injection experiments using real failure statistics from the field {(in Section~\ref{sec:Results and Observations}) show}  that this feature of PMDS code can dramatically decrease the number of data loss events compared to RAID5, at a performance overhead and negligible space overhead.

\subsection{Statistical Fault Injection Environment}
\label{sec:fault injection environment}
  
  \begin{figure}
    \centering
 
        \includegraphics[width=0.48\textwidth]{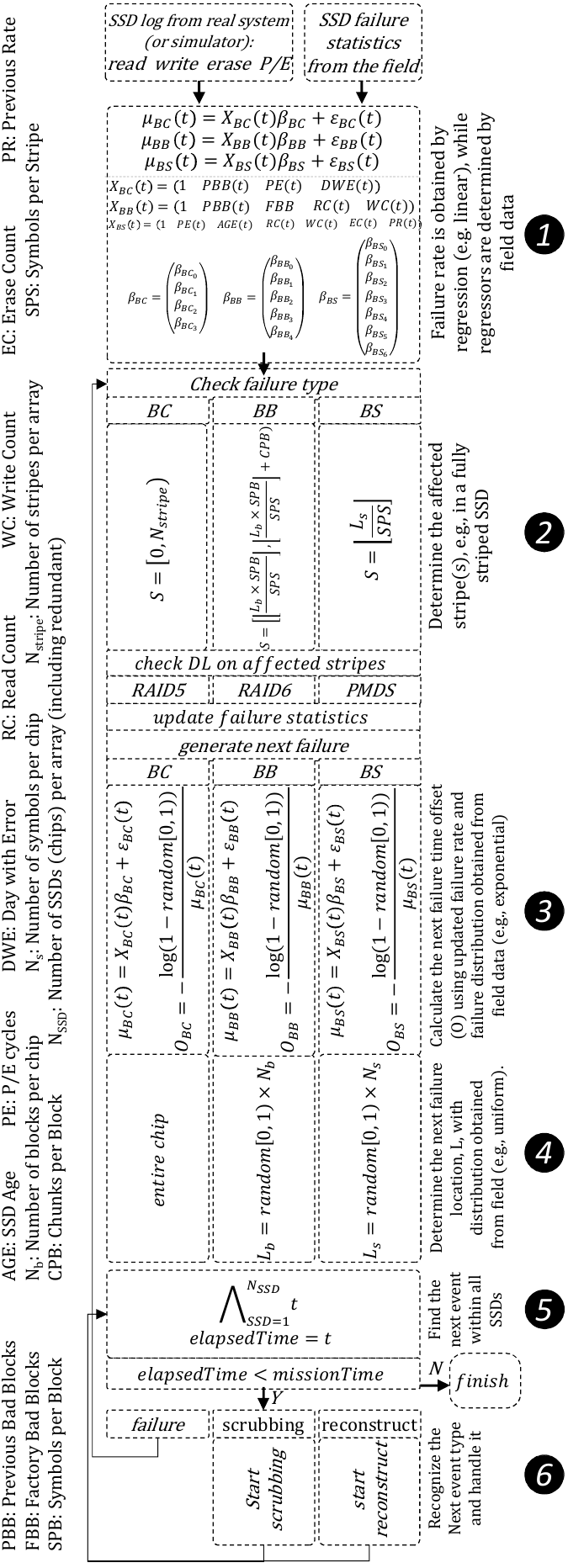}
    \caption{Statistical fault injection {environment.}}
    \label{fig:montecarlo}

\end{figure}
  
Fault injection can be implemented by injecting faults on SSD simulator (such as DiskSim~\cite{bucy2008disksim}) in the runtime, as previous work~\cite{li2016analysis} does. 
{That} approach, however, is very {time-consuming} as SSD simulators have much complexities and are very slow. 
We take another approach that extracts SSD usage log, including the number of reads, writes, erases, and P/E cycles from a real system (or simulator), and {post-processes} this information to perform fault injection experiments.   
This approach has two advantages: \textbf{a)} It is {very} fast.
\textbf{b)} We can use SSD usage logs from the real systems, rather than simulators, to obtain more realistic results. 
{Accordingly, our failure model has two phases: a) \emph{Capturing SSD Logs} and b) \emph{Statistical Fault Injection} in respect with SSD failure statistics from the field and SSD usage logs obtained by our real-system run. }

\subsubsection{{Capturing SSD Logs}}
{In the phase of capturing SSD log, we have different benchmarks running on the desired array configuration. What we capture from benchmark running is the number of P/E cycles as well as write/read accesses. In specific, P/E cycles are calculated by capturing \emph{Wear Leveling Count} parameter from S.M.A.R.T~\cite{rothberg2005disk}, before and after the benchmark run, following the instructions of SSD vendors. The number of writes confirmed to the SSD is also calculated by capturing \emph{Total LBAs Written} from S.M.A.R.T~\cite{rothberg2005disk}, before and after the benchmark run. 
For each individual SSD, we also extract the full details of read/write request type, destination, size, and issue time using Blktrace~\cite{blktrace}. }

\subsubsection{{Statistical Fault Injection}}

Fig.~\ref{fig:montecarlo} shows the flowchart of our fault injection framework.
We use the field data of SSD failure statistics from Schreoder et al.~\cite{schroeder2016flash}, alongside SSD operation log to dynamically evaluate the rate of bad chip, bad block, and bad symbol, per each individual chip. 
Schreoder et al.~\cite{schroeder2016flash} show that failure (bad chip, bad block, and bad symbol) rate at time $t$ highly correlates with parameters such as \emph{Read Count} ($RC$), \emph{Write Count} (WC), \emph{Erase Count} ($EC$), \emph{Previous Bad Blocks} ($PBB$), \emph{Previous bit error Rate} ($PR$), \emph{Days With Error} ($DWE$), and number of \emph{Factory Bad Blocks} {($FBB$).}\footnote{Bad blocks already exist on a brand-new SSD chip~\cite{schroeder2016flash}.} Hence, the failure rates (bad chip, bad block{, and} bad symbol rates) are evaluated by regression ({e.g.,} linear regression) from the mentioned factors, while the \emph{Parameter Vector}, $\beta$, should be determined by the field data.
Accordingly, the rate of bad chip, $\mu_{BC}(t)$, is calculated as shown in Equation~\ref{equ:regression} (considering linear regression).
\begin{equation}
\label{equ:regression}
\begin{split}
\mu_{BC}(t)=X_{BC}(t) \times \beta_{BC} + \epsilon_{BC}(t) \\
X_{BC}(t) = 
\begin{bmatrix}
1 & PBB(t) & PE(t) & DWE(t)
\end{bmatrix}
\\
\beta_{BC} = 
  \begin{bmatrix}
\beta_1 \\
\beta_2 \\
\beta_3 \\
\beta_4
   \end{bmatrix}
\end{split}
\end{equation} 

Where $X_{BC}(t)$ is the vector of regressors determined by field data, $\beta_{BC}(t)$ is the parameter vector (also determined by field data), and $\epsilon_{BC}(t)$ is the error vector. 
$\mu_{BB}(t)$ and $\mu_{BS}(t)$ are also calculated by the similar equations shown in Fig.~\ref{fig:montecarlo}.
Note that Schroeder et al.~\cite{schroeder2016flash} have reported a limited failure data including RBER as a function of $EC$, percentage of drives with bad blocks, median number of bad blocks, mean number of bad blocks, and percentage of drives with bad chips. 
We determine the failure rates by using the mentioned data. Fig.~\ref{fig:ssd failure statistics} and Table~\ref{tab:field-failure-statistics} summarize the employed failure statistics by Schroeder et al.~\cite{schroeder2016flash}.

{In the fault injection phase, the desired array of SSDs is constructed and for each individual SSD, the P/E cycles and accesses are imported from SSD logs captured by benchmark run. For each SSD, the rate of $BS$ ($\mu_{BS}(t)$) is updated in specific time intervals, regarding P/E cycles at time $t$, following the field data appeared in Table~\ref{tab:field-failure-statistics} and Fig.~\ref{fig:ssd failure statistics}. 
The other two parameters of $BB$ and $BC$ are also considered for each individual SSD device. About $BC$, we have the rate of drives with bad chip (in a four-years mission) from the field~\cite{schroeder2016flash}. The available detail on how bad chips correlate bad blocks is limited to the fact that 2/3 of all bad chips appear in those chips that have more than 5\% of their blocks failed. We consider this correlation in our experiments by creating a pool of SSDs at the beginning of experiments, following the failure statistics reported by the field data.}

\textbf{{Create SSD Pool:}}
{We build a pool of SSDs (in our experiments, 10,000 SSDs) regarding the failure attributes reported by~\cite{schroeder2016flash}. In the next step, we construct the disk array by randomly selecting \emph{n} SSDs (e.g., 8 SSDs for an array of RAID5(7+1)) from the SSD pool. Following we describe how bad chip and bad block is considered in the failure model.
}

\textbf{{Bad Chip:}}
{At the start of fault injection experiments, the SSD pool is created in a way the $BC$ and $BB$ statistics conforms the field data reported by~\cite{schroeder2016flash} (detailed results on how the $BB$ and $BC$ of SSDs in the failure model statistically conforms the field data are appeared in Section~\ref{sec:validating-regression-model}). When the SSD pool is created, some chips are marked as to be failed within mission time. 
In the SSD array, constructed by randomly choosing \emph{n} out of 10,000 SSDs in the pool, if a chip is marked to encounter bad chip, it is failed within mission time. No data is provided by~\cite{schroeder2016flash} about the time distribution of BC, so we consider exponential distribution, following the conventional assumption on the time to failure of semiconductors. Following we describe how the correlation between $BC$ and $BB$ is considered in the failure mode.}

\textbf{{Bad Block:}}
{The distribution of the number of mission-time bad blocks in the SSD population is not reported by field data. The field data only reports that the number of \emph{factory} bad blocks is close to \emph{Normal} distribution in the population of SSDs under study~\cite{schroeder2016flash}. Hence, in creating the SSD pool we consider the number of mission-time bad blocks also follows the normal distribution. 
From the field data, we also have the percentage of drives with bad blocks, median number of bad blocks for drives having bad block, and mean number of bad blocks for drives having bad block. We create the SSD pool to conform the mentioned statistics obtained from the field, as shown in Section~\ref{sec:validating-regression-model}.
The field data also reports a correlation between bad chip and bad block~\cite{schroeder2016flash}. Based on the field results, 2/3 of all bad chips happen in those chips that have more than 5\% of all their blocks failed. We consider this correlation between $BB$ and $BC$ in creating the SSD pool, as shown in Section~\ref{sec:validating-regression-model}. 
The last correlation reported by the field study is the correlation of $BB$ with previous $BB$~\cite{schroeder2016flash}. The field study reports the median number of bad blocks a drive will experience within mission time, as a function of number of bad blocks already experienced. We also consider this correlation in creating the SSD pool by increasing $BB$ probability in those chips that have experienced a specific threshold of $BB$, as verified in Section~\ref{sec:validating-regression-model}.
Regarding the mentioned statistical attributes, we determine the occurrence of bad chip, as well as the number of bad blocks for each SSD in the SSD pool.  }

\textbf{{Constructing SSD Array and Starting Fault Injection:}}
{In the next step, we construct the SSD array by randomly choosing $n$ out of 10,000 SSDs from the pool. For each SSD, we also have P/E cycles as well as accesses from benchmark run on the real system. 
As the fault injection time, $t$, passes, for each SSD the $RBER$ is updated regarding P/E cycles at time $t$, following the data appeared in~\cite{schroeder2016flash} (Fig.~\ref{fig:ssd failure statistics}). 
For time intervals of one hour, we estimate the number of $BS$ by multiplying $RBER$ to the number of accessed bits.
No information on the spatial characteristics of BS is reported by the field data, so we consider uniform distribution for BS.
The time distribution of $BS$, $BB$, and $BC$ is not clarified by the field data, so we consider exponential distribution for time to failure, following the conventional assumption on the time to failure of semiconductor devices. }

At the beginning of fault-injection, the simulator is initiated by the first bad chip, bad block and bad symbol for each chip, and the next failure (the failure having minimum $t$) is issued to the failure handling queue. 
Thereafter, the simulator recognizes the failure type and determines the affected sectors, regarding the failure type and location. 
As we consider a fully-striped architecture for SSD array, in the case of bad {chips} all stripes are affected (stripes 0 to $N_{stripe}$, as shown in Fig.~\ref{fig:montecarlo}). 
In the case of bad blocks, the number of affected stripes is equal to \emph{Chunks Per Block} (CPB) that depends on the array architecture, including block size, stripe size, and number of devices. 
The index of the first affected stripe is $\lfloor \frac{L_b \times SPB}{SPS} \rfloor$, where $L_b$ is the location (index) of affected block, $SPB$ is the number of symbols per block, and $SPS$ is the number of symbols per stripe.
In the case of bad symbols, a single stripe is affected which index is $\lfloor \frac{L_s}{SPS} \rfloor$, where $L_s$ is the location (index) of the affected symbol and $SPS$ is the number of symbols per stripe.

Afterwards, the simulator needs to check if any of the previous errors in the affected sectors have already {removed.}\footnote{Previous bad blocks and bad sectors are possibly detected after a read error, write error, or erase error, and removed by reallocating in the case of bad blocks, and rewriting in the case of bad symbol. Scrubbing and SSD reconstruct also remove the errors, but these tasks are handled in another procedures.}
In the next step, {based} on the analysis described in Section~\ref{sec:analysis erasure codes}, the simulator checks data loss (ADL, BDL, and SDL) on the affected stripes, regarding the employed erasure code (RAID5, RAID6, or PMDS) and updates the failure statistics. 

After handling a failure, it is needed to generate the next failure of that type. For example, after handling a bad symbol on chip $c$, it is needed to generate the next bad {symbol} incidence for chip $c$. 
To this end, the simulator determines the next failure time offset, $O$, using the dynamically evaluated failure rate, $\mu$. 
Considering time-to-failure follows exponential distribution, the time offset of the next bad chip, $O_{BC}$ is recognized as {shown} in Equation~\ref{equ:next-failure-time}.
\begin{equation}
\label{equ:next-failure-time}
\begin{split}
O_{BC} = \frac{log(1-random[0,1))}{\mu_{BC}(t)}
\end{split}
\end{equation} 

Where $random[0,1)$ is a uniformly generated random number between 0 and 1, and $\mu_{BC}(t)$ is the rate of bad chip at time $t$, dynamically evaluated using Equation~\ref{equ:regression}.
The time offset of the next bad block, $O_{BB}$, as well as the time offset of the next bad symbol, $O_{BS}$, are also calculated with the similar equations, as shown in {Fig.~\ref{fig:montecarlo}.}
Thereafter, the simulator determines the location of failure (in the case of bad symbol and bad block), regarding the number of symbols and number of blocks per chip, with predefined distribution obtained from the field (e.g., uniform distribution within a single chip). 
Accordingly, the location of the next bad symbol event, $L_s$ is determined as shown in Equation~\ref{equ:next-failure-location}.
\begin{equation}
\label{equ:next-failure-location}
\begin{split}
L_s = random[0,1) \times N_s
\end{split}
\end{equation} 

Where $N_s$ is the number of symbols per array.
The location of the next bad block event, $L_b$, is determined by a similar equation, as shown in Fig.~\ref{fig:montecarlo}.
In the next step, the next event (failure, scrubbing, or reconstruct) within all chips is issued for handling, and the simulation time is set to the next event time (the event with the minimum time offset, as shown in Fig.~\ref{fig:montecarlo}). If the total mission time is already passed, the simulation finishes. Otherwise, in the case of \emph{reconstruct}, the {reconstruction} is {starting}, the possible data loss detected in the {reconstruction} process is collected, and the replaced SSD statistics is initiated. 
In the case of \emph{scrubbing}, the possible data loss detected in the scrubbing process is collected and the correctable errors is removed. 
{Finally,} in the case the next event is \emph{failure} the simulator turns back to the state of checking the failure type.

\section{Results and Observations}
\label{sec:Results and Observations}
\begin{figure}
	\centering
	
	\includegraphics[width=0.35\textwidth]{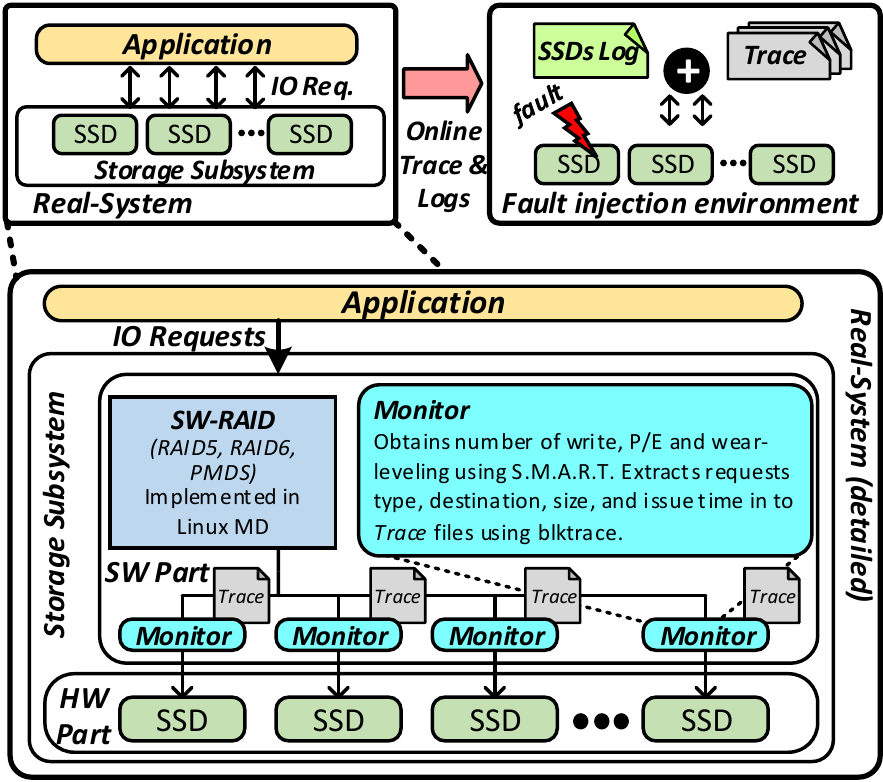}
	\caption{{The structure of test platform used in experiments.}}
	
	\label{fig:test-platform}
	
\end{figure}

In this {section,} we evaluate the reliability and performance of RAID5, RAID6, and PMDS array configurations using the test platform depicted in Fig. \ref{fig:test-platform}. {We examine realistic application workloads on the SSD arrays under a real platform and track the block layer {I/O} traces as well as SSD usage statistics provided by S.M.A.R.T. }
 The performance of different array configurations is collected from workload execution on the real platform, while the array reliability is obtained from our fault injection framework (presented in Section~\ref{sec:fault injection environment}) by {post-processing} the SSD usage logs. In the following we first elaborate the details of test platform and examined SSDs. Afterwards, we provide the {experimental} results.

\subsection{Experimental Setup}
\label{sec:experimental setup}

 Our test platform is {composed of} \emph{real-system} and \emph{fault injection environment}. 
In the real-system part of the platform, we use an open-source software RAID controller, MD driver in Linux kernel version 3.10.0-327 running on \emph{CentOS 7} operating system, and arrays of Samsung 850 Pro SSDs to obtain the effect of erasure codes on the performance of SSD arrays, and capture the SSD usage logs. 
Table \ref{table:hardware} provides the details of hardware and software stack in real-system. 
{This platform} is responsible for array performance evaluation and collecting SSD usage logs. 
Afterwards, the array reliability is evaluated using the fault injection environment by {post-processing} the I/O traces and SSD usage logs collected from the real-system run (as detailed in Section~\ref{sec:fault injection environment}).
The statistical fault injection environment is developed from scratch in C++.
{The pseudo-code of the major functions of fault injection implementation is shown in Algorithm~\ref{pseudocode} in Appendix \ref{sec:app-alg}.} 
{The supplementary function definitions are also shown in Algorithm~\ref{pseudocode2} in Appendix \ref{sec:app-alg}.}

\begin{table}[]
	\centering
	\caption{Hardware and software stack of the real-system part of our test platform, responsible for examining the performance of SSD array configurations and collecting SSD usage logs.}
	\vspace{-0.1cm}
	\label{table:hardware}
	\begin{tabular}{|l|l|}
		\hline
		\multicolumn{2}{|l|}{\textit{\textbf{Software}}}                 \\ \hline\hline
		OS                 & CentOS 7                                    \\ \hline
		Kernel             & 3.10.0-327                                  \\ \hline
		SW RAID controller & \textit{Multiple Device (MD) driver}        \\ \hline\hline
		\multicolumn{2}{|l|}{\textit{\textbf{Hardware}}}                 \\ \hline\hline
		Under test SSDs    & 8x Samsung 850 Pro, 512GB, SATA             \\ \hline
		RAM                & 8GB from Hynix Semiconductor                \\ \hline
		CPU                & 16 core Intel (R) Xeon (R) E5-2620 @ 2.1GHz \\ \hline
		Motherboard        & Supermicro X10DRL-i                         \\ \hline
	\end{tabular}
\end{table}

\subsubsection{Workloads}
{The experiments are conducted using both synthetic benchmarks and realistic applications.} For the synthetic {experiments,} we employ FIO tool~\cite{fio} and run \emph{Random Read} ($RR$), \emph{Random Write} ($RW$), \emph{Sequential Read} ($SR$), \emph{Sequential Write} ($SW$){, and} mixture of random read/write requests ($Mixed$) workloads. {We also employ \emph{Filebench}~\cite{tarasov2016filebench} in order to commit realistic application I/O requests to the disk subsystem.} We run various workloads, including \emph{Webserver}, \emph{Fileserver}, \emph{Varmail}, \emph{Copyfiles}, \emph{Mongo}, and \emph{Video server} from \emph{Filebench} framework.
 In the following, we explain the characteristics of examined benchmarks. 
 \begin{itemize}
 	\item 
 	\textbf{\emph{FIO}}  is a powerful synthetic benchmark, capable of generating synthetic workloads with customized access pattern, request type, request size, locality of accesses, and \emph{Working Set Size} (WSS). Using FIO, we examine five representative synthetic workloads as detailed in Table~\ref{table:fio_workloads}.
 	
	\item 
	\textbf{\emph{Filebench}} is a benchmarking tool that works on the filesystem level and can generate a big spectrum of application workloads. In our {experiments,} we employ six representative workloads including \emph{Webserver}, \emph{Varmail}, \emph{Webproxy}, \emph{Mongo}, \emph{Video server}, and \emph{Fileserver}. 
The \emph{Webserver} workload creates millions of files with mean size equal to 64KB and the maximum request size equal to 1MB where more than 100 threads have access to the files at the same time.
Similarly, the \emph{Varmail} workload creates files with mean size equal to 16KB with a maximum request size equal to 1MB, but the number of threads is equal to 16 which is significantly less than \emph{Webserver}.

The \emph{Fileserver} workload creates more than 600,000 files with mean size equal to 128KB and the maximum request size equal to 1MB while more than 50 threads have simultaneous access to the files.
The \emph{Mongo} workload which simulates the MongoDB {I/O} {requests} creates 100,000 files with mean size equal to 512KB and the maximum request size equal to 1MB while only one thread commits I/O request to the files.
The last workload is \emph{Video server} that creates files with the size of 10GB in average, including 6 active and 7 inactive videos. The event rate of this workload is equal to 96 and the average I/O size is equal to 256KB which only includes read requests.
{We should note that Filebench experiments are performed with both enabled and disabled buffer cache. We mainly disable buffer cache to evaluate the performance of the disk subsystem (i.e., array of SSDs) and to remove the impact of filesystem level cache on the {I/O} requests. Our initial experiments reveal that enabled buffer cache reduces the number of committed writes into the SSD array by 2X (on average). We observe high level of reduction in two types of workloads: a) read-intensive ones with few number of write operations such as videoserver and fileserver and b) write-intensive workloads with large number of Write-After-Write (WAW) sequences with small reuse distance. The former type of workloads has a limited number of write accesses that are mostly handled by DRAM in the presence of buffer cache while in the latter type of workloads, the write accesses and further modifications (second write operation on the same address) are mainly buffered in the buffer cache. Hence, by enabling the buffer cache we experience considerably fewer write operations on the disk subsystem (\emph{varmail}, \emph{randwrite}, and \emph{randreadwrite} workloads best fit in this group). For the rest of workloads such as \emph{mongo}, \emph{copyfiles}, and \emph{webserver}, since these workloads almost include equal number of read and write operations (with sequential pattern in \emph{copyfiles} and random pattern in \emph{mongo} and \emph{webserver}), buffer cache is of less improving effect and helps to reduce the write operations respectively by 1.07X, 1.6X, and 1.5X in \emph{mongo}, \emph{copyfiles}, and \emph{webserver}.}
 \end{itemize}

\begin{table}[!h]
	\centering
	\caption{Parameters of the running workloads with FIO.}
	\vspace{-0.2cm}
	\label{table:fio_workloads}
	\scriptsize
	\begin{adjustbox}{width=0.5\textwidth,totalheight=\textheight,keepaspectratio}
	\begin{tabular}{|c||c|c|c|c|c|c|c|}
		\hline
		\textbf{Workload} & \begin{tabular}[c]{@{}c@{}}
			\textbf{Req.} \\
			\textbf{Size}
		\end{tabular} & \begin{tabular}[c]{@{}c@{}}\textbf{Req.}\\ \textbf{Type}\end{tabular}               & \begin{tabular}[c]{@{}c@{}}\textbf{Access}\\ \textbf{Pattern}\end{tabular}                   & \begin{tabular}[c]{@{}c@{}}{\textbf{I/O}}\\ \textbf{depth}\end{tabular} & \textbf{Threads} & \begin{tabular}[c]{@{}c@{}}{\textbf{I/O}}\\ \textbf{Engine}\end{tabular} \\ \hline\hline
		SR       & 4MB                                                 & Read                                                              & Sequential                                                                 & 16                                                 & 1       & libaio                                              \\ \hline
		SW       & 4MB                                                 & Write                                                             & Sequential                                                                 & 16                                                 & 1       & libaio                                              \\ \hline
		RR       & 4KB                                                 & Read                                                              & \begin{tabular}[c]{@{}c@{}}Random\\ (distribution:\\ zipf:1.2)\end{tabular} & 16                                                 & 16      & libaio                                              \\ \hline
		RW       & 4KB                                                 & Write                                                             & \begin{tabular}[c]{@{}c@{}}Random\\ (distribution:\\ zipf:1.2)\end{tabular} & 16                                                 & 16      & libaio                                              \\ \hline
		Mixed    & 4KB                                                 & \begin{tabular}[c]{@{}c@{}}Read/Write\\ (read: 70\%)\end{tabular} & \begin{tabular}[c]{@{}c@{}}Random\\ (distribution:\\ zipf:1.2)\end{tabular} & 16                                                 & 16      & libaio                                              \\ \hline
	\end{tabular}
\end{adjustbox}
\end{table}

 {The array performance for each experiment is collected from the output of the FIO and Filebench (reported in Appendix \ref{sec:perf}). We validate the performance output of the benchmarks using \emph{iostat} tool from \emph{sysstat} package of the Linux.  
 The block-layer log of logical accesses to individual SSDs and the SSD array (i.e., virtual disk) is collected by using \emph{blktrace}~\cite{blktrace}. \emph{blktrace} is a comprehensive {I/O} tracing tool of Linux kernel that monitors the {I/O} requests committed and responded by the SSDs and virtual disk.
 The exact number of writes to each SSD and the number of wear leveling and Program/Erase (P/E) operations performed on each SSD is also obtained by using S.M.A.R.T~\cite{rothberg2005disk}. These statistics are used later in the fault injection process to dynamically evaluate the failure rates 
 and also reported as endurance results in Appendix \ref{sec:endurance}.
Table~\ref{tab:ssd array configuration} shows the basic configuration of examined SSD array. In some experiments, we have modified some parameters that are noted in case. 
Note that we examine different erasure codes under fixed physical capacity to have a fair performance comparison.
Using the collected statistics, we conduct fault injection experiments for 4 years mission time, while 1000 experiments are conducted per configuration.}

\begin{table}
\centering
\caption{{SSD array basic configuration.}}
    \vspace{-0.3cm}
\begin{center}
	\scriptsize
    \begin{tabular}{ | c | c ||c|c|}
    \hline
     \textbf{Parameter} & \textbf{value}  & \textbf{Parameter} & \textbf{value} \\ \hline\hline
     SSD Elements & 8 & SSD Page Size & 4 KB \\ \hline
SSD Page Per Block & 64 & SSD Planes Per Element &8 \\ \hline
 SSD Block Per Element &$16,384$ & SSD Stripe Size& 128 KB \\ \hline
SSD Size & $512~GB$ & SSD Blocks &$131,072$  \\ \hline
 Array Devices & 8  & Chunk Pages & 4  \\ \hline 
     
    \end{tabular}
\end{center}
\label{tab:ssd array configuration}
\end{table}

\subsection{Fault Injection Parameters}
\label{sec:fault-injection-parameters}
\begin{figure}
    \centering
    
        \includegraphics[width=0.5\textwidth]{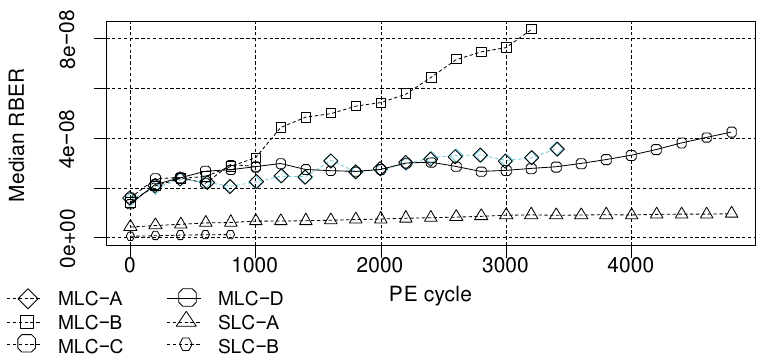}
   \vspace{-0.3cm}    
 
    \caption{{Summary of} SSD RBER statistics as a function of P/E cycles in {4 years of} mission time~\cite{schroeder2016flash}.}

    \label{fig:ssd failure statistics}
\end{figure}

We used SSD failure statistics by~\cite{schroeder2016flash} that investigates the failure of six SSD chip models, including four chip with MLC technology and two with SLC technology. 
We use the median RBER reported by~\cite{schroeder2016flash} (Fig.~\ref{fig:ssd failure statistics}) as a function of {P/E} cycles for each SSD model.
Within fault injection experiments, RBER for each chip is dynamically determined regarding the number of P/E cycles at time $t$, obtained from SSD usage log.
Using RBER, the \emph{Bit Error Rate} (BER) of each SSD device is dynamically evaluated regarding the number of accessed bits, obtained from SSD usage logs, as follows:
\begin{equation}
\label{equ:ber}
\begin{split}
BER = RBER \times number~of~accessed~bits(\Delta t)
\end{split}
\end{equation} 

Where $\Delta t$ is the time interval the BER is evaluated for. We conduct our experiments by considering $\Delta t = 1hour$.
We also use the percentage of drives with bad chips and bad blocks reported for each SSD model in a {four} year mission~\cite{schroeder2016flash}, as well as the mean and median number of bad blocks for each SSD model (appeared in Table~\ref{tab:field-failure-statistics})~\cite{schroeder2016flash}. 
Regarding these statistics and a restriction reported by~\cite{schroeder2016flash} that the chip is considered failed when 5\% of its blocks are failed (happened in $2/3$ of all bad chip cases), we determine bad chip time and bad block rate for each SSD chip.

\begin{table}
\centering
\caption{ {{Summary of} SSD Bad chip and bad block statistics in {four years of} mission time~\cite{schroeder2016flash}.}}
    \vspace{-0.3cm}
\begin{center}
	\scriptsize
\begin{adjustbox}{width=0.5\textwidth,totalheight=\textheight,keepaspectratio}
    \begin{tabular}{ | c || c | c | c | c | c | c |}
    \hline
 \textbf{Model Name}    &\textbf{MLC-A} &\textbf{MLC-B} & \textbf{MLC-C}& \textbf{MLC-D}& \textbf{SLC-A}& \textbf{SLC-B}\\ \hline\hline
\% Drives w/ Bad Chips&  5.6&6.5 &6.6 &4.2 &3.8&2.3 \\ \hline
\% Drives w/ Bad Blocks&31.1 &79.3 & 30.7&32.4 &39.0 &64.6 \\ \hline
Median \# Bad Blocks &2 & 3& 2& 3& 2& 2\\ \hline
Mean \# Bad Blocks & 772&578 &555 &312 &584 &570 \\ \hline
     
    \end{tabular}
  \end{adjustbox}
\end{center}
\label{tab:field-failure-statistics}
\end{table}

\subsection{{Validating Regression Model}}
\label{sec:validating-regression-model}
{Table~\ref{tab:regression-validation} shows the SSD failure statistics of 10,000 drives in our regression model.
Comparing the output of the regression model with field data results, the regression model fully matches the field data~\cite{schroeder2016flash} in the following parameters:}
\begin{itemize}
	\item
	{Drives with bad chip}
	\item 
	{Drives with bad blocks}
	\item 
	{Median number of bad blocks} 
\end{itemize}

\begin{table}[]
	\centering
	\caption{{Failure statistics of 10,000 SSD drives in the regression model.}}
	\vspace{-0.3cm}
	\begin{center}
		\scriptsize
		\begin{adjustbox}{width=0.5\textwidth,totalheight=\textheight,keepaspectratio}
	\begin{tabular}{|c|c|c|c|c|c|c|}
		\hline
		& \textbf{MLCA} & \textbf{MLCB} & \textbf{MLCC} & \textbf{MLCD} & \textbf{SLCA} & \textbf{SLCB} \\ \hline
		\textbf{Drives W Bad Blocks}                                                                                                             & 3100          & 7930          & 3070          & 3240          & 3900          & 6460          \\ \hline
		\textbf{Median \# Bad Blocks}                                                                                                            & 2             & 3             & 2             & 3             & 2             & 2             \\ \hline
		\textbf{Mean \# Bad Blocks}                                                                                                              & 769           & 555.08        & 557.13        & 347.54        & 551.68        & 559.88        \\ \hline
		\textbf{Drives W Bad Chips}                                                                                                              & 560           & 650           & 660           & 420           & 380           & 230           \\ \hline
		\textbf{\begin{tabular}[c]{@{}c@{}}Drives with BC \\ and BB in more than\\  5\% of all blocks\end{tabular}}                              & 375           & 435           & 442           & 281           & 254           & 154           \\ \hline
		\textbf{\begin{tabular}[c]{@{}c@{}}Rate of drives with\\ BC and BB in more \\ than 5\% of all blocks\\ over drives with BC\end{tabular}} & 0.67          & 0.67          & 0.67          & 0.67          & 0.67          & 0.67          \\ \hline
		\textbf{\begin{tabular}[c]{@{}c@{}}Drives with BC but \\ no BB (not reported \\ by the field data)\end{tabular}}                         & 141           & 48            & 157           & 95            & 76            & 25            \\ \hline
	\end{tabular}
  \end{adjustbox}
\end{center}
\vspace{-0.3cm}
\label{tab:regression-validation}
\end{table}

\begin{figure}
	\centering
	
	\includegraphics[width=0.4\textwidth]{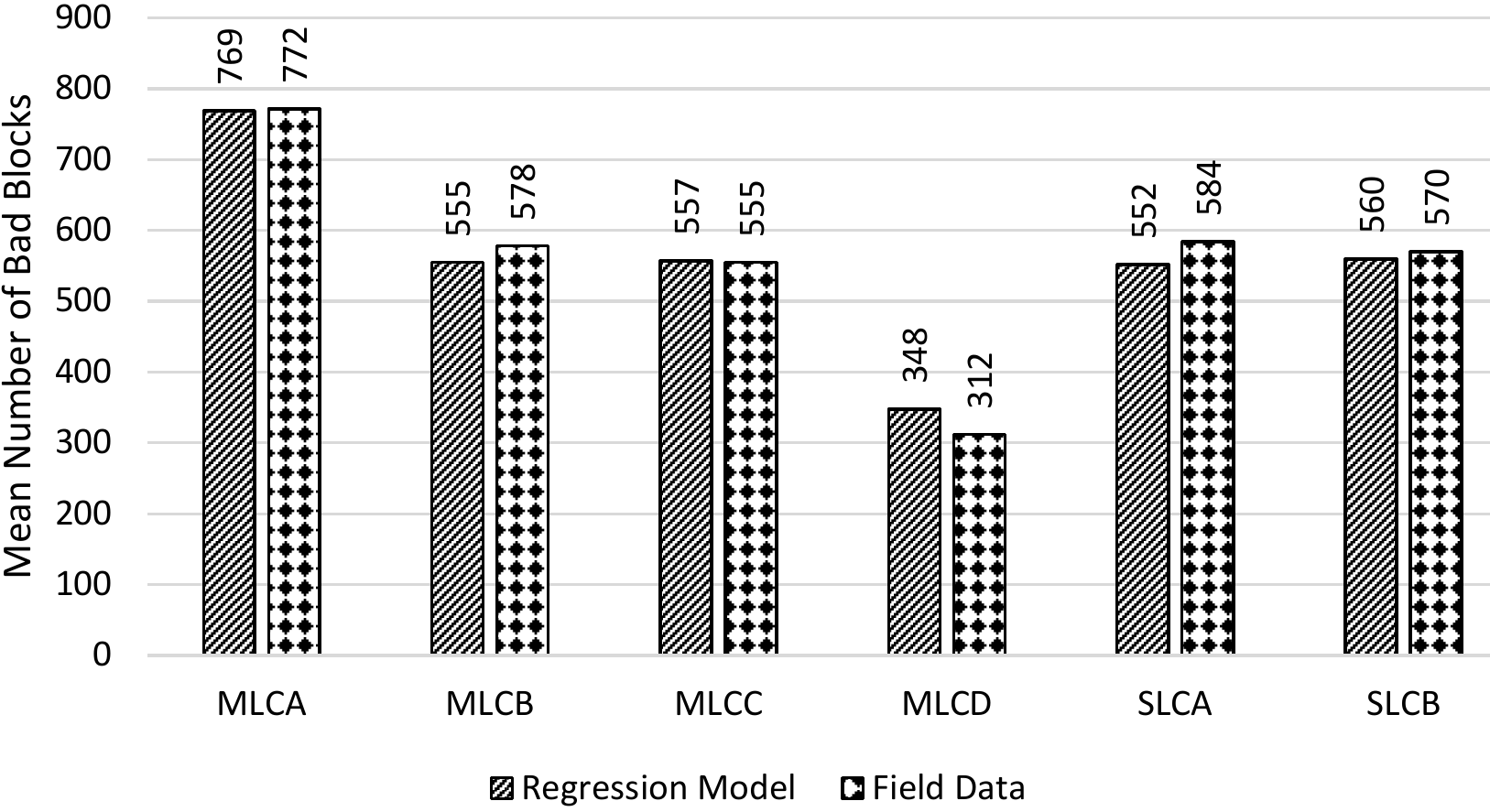}
	
	\caption{{Mean number of bad blocks reported by field data~\cite{schroeder2016flash} and regression model.}}
	
	\label{fig:regression-model-output}
\end{figure}

{Mean number of bad blocks has a maximum 11\% error (for MLCD) compared to the field data, as shown in Fig.~\ref{fig:regression-model-output}. Another constraint obtained by field data is the correlation between bad chip and bad block. Indeed, 2/3 of all bad chips happen in those chips that have more than 5\% of all their blocks failed. Table~\ref{tab:regression-validation} reports the rate of drives with $BC$ and $BB$ in more than 5\% of all blocks, over drives with BC, showing that our regression model is loyal to this correlation.}

{Another correlation reported for bad blocks is the median number of bad blocks a drive will experience within mission time, as a function of number of bad blocks already experienced. The field data shows a very steep increase in median number of $BB$ when the chip experiences more than one (in the case of MLC) and more than three (in the case of SLC) bad blocks. In specific in the case of MLC, the median number of bad blocks jumps to 200 after the second bad block is experienced. We consider this constraint in our regression model by increasing the rate of $BB$ in those chips that have already experienced two bad blocks (four in the case of SLCs). However, as the field data does not distinguish the mentioned statistics for different MLC/SLC models (as appeared in Figure 8 of~\cite{schroeder2016flash}), we employ the average values reported for SLCs and MLCs. Our empirical results show that the regression model is loyal to this constraint, as shown in table~\ref{tab:regression-median-validation}. Table~\ref{tab:regression-median-validation} shows the median number of bad blocks as a function of previous number of bad blocks experienced in the regression model output and field data (the average value reported for SLC and MLC is reported).}

\begin{table}[]
	\centering
	\caption{{Median number of bad blocks as a function of previous number of bad blocks experienced in the regression model and field data~\cite{schroeder2016flash}.}}
	\vspace{-0.3cm}
	\begin{center}
		\scriptsize
		\begin{adjustbox}{width=0.5\textwidth,totalheight=\textheight,keepaspectratio}
			\begin{tabular}{|c|c|c|c|c|c|c|c|c|}
				\hline
				\textbf{\begin{tabular}[c]{@{}c@{}}Previous Num.\\ of Bad Blocks\end{tabular}} & \multicolumn{2}{c|}{\textbf{2}}                                                                                               & \multicolumn{2}{c|}{\textbf{3}}                                                                                               & \multicolumn{2}{c|}{\textbf{4}}                                                                                               & \multicolumn{2}{c|}{\textbf{5}}                                                                                               \\ \hline
				& \textbf{\begin{tabular}[c]{@{}c@{}}Reg.\\ Model\end{tabular}} & \textbf{\begin{tabular}[c]{@{}c@{}}Field\\ Data\end{tabular}} & \textbf{\begin{tabular}[c]{@{}c@{}}Reg.\\ Model\end{tabular}} & \textbf{\begin{tabular}[c]{@{}c@{}}Field\\ Data\end{tabular}} & \textbf{\begin{tabular}[c]{@{}c@{}}Reg.\\ Model\end{tabular}} & \textbf{\begin{tabular}[c]{@{}c@{}}Field\\ Data\end{tabular}} & \textbf{\begin{tabular}[c]{@{}c@{}}Reg.\\ Model\end{tabular}} & \textbf{\begin{tabular}[c]{@{}c@{}}Field\\ Data\end{tabular}} \\ \hline
				\textbf{MLC}                                                                     & 143                                                           & 143                                                           & 155                                                           & 151                                                           & 159                                                           & 158                                                           & 183                                                           & 183                                                           \\ \hline
				\textbf{SLC}                                                                     & 5                                                             & 5                                                             & 20                                                            & 20                                                            & 43                                                            & 45                                                            & 77                                                            & 75                                                            \\ \hline
			\end{tabular}
		\end{adjustbox}
	\end{center}
	\label{tab:regression-median-validation}
\end{table}

\subsection{Data Loss Breakdown}
\begin{figure}
    \centering
 
        \includegraphics[width=0.5\textwidth]{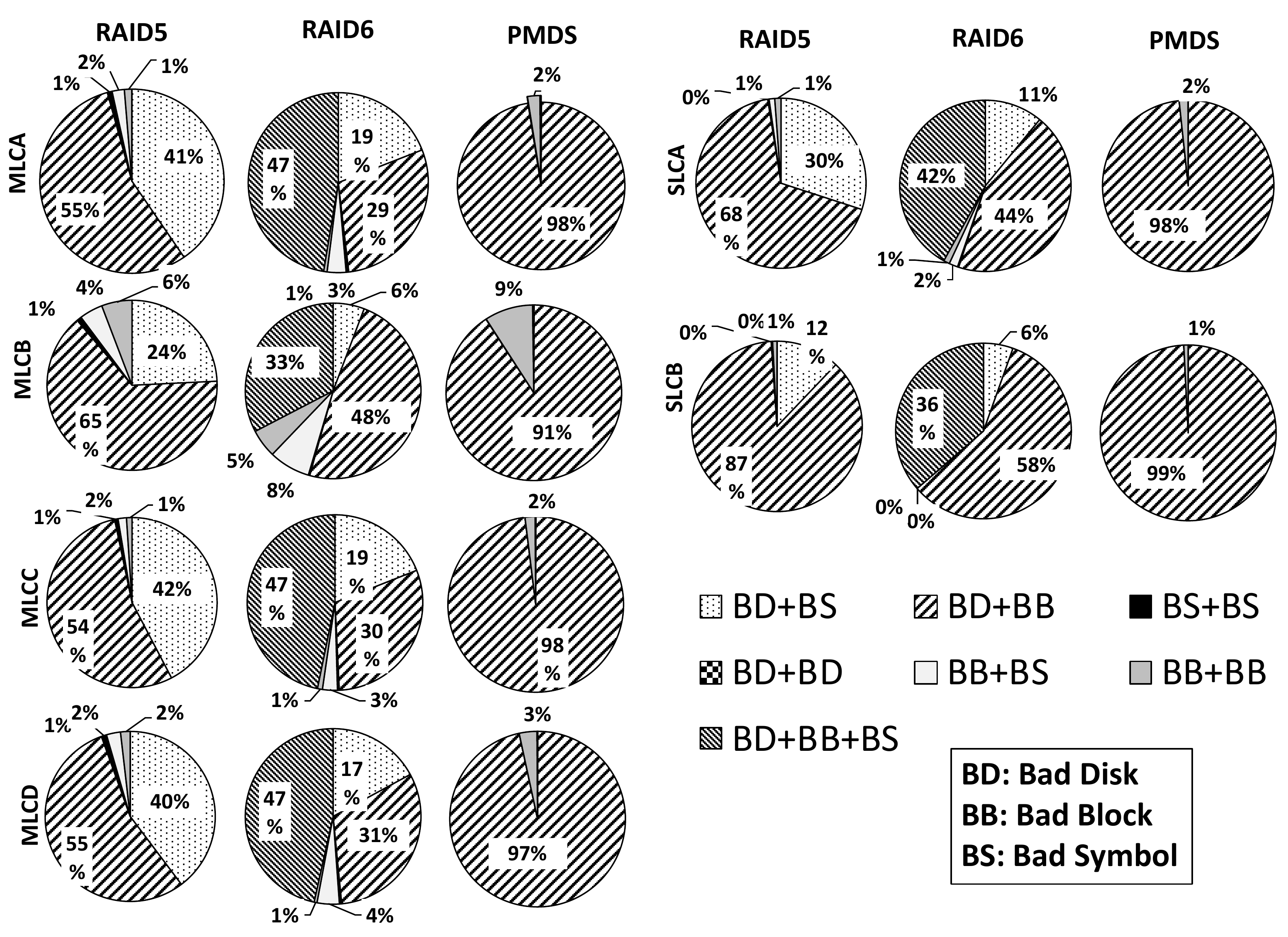}
    \caption{{Failure breakdown for different erasure codes and SSD types (TTS = 10,000h, TTR = 10h)}}
    \label{fig:failure-breakdown}

\end{figure}

{Fig.~\ref{fig:failure-breakdown} shows data loss root cause breakdown for different erasure codes and SSD types. 
The results are reported for aggregate of lost stripes in 12 examined workloads (including both synthetic and realistic application workloads), assuming \emph{Time to Scrub} (TTS)\footnote{TTS is the expected time between two array scrubbing processes.} equal to 10,000 Hours and \emph{Time to Recover} (TTR)\footnote{In the case of SSD failures, TTR is the expected time of device recovery process.} equal to 10 Hours.
The figure shows how different combinations of failures, including Bad Chip, Bad Block, and Bad Symbol, contribute to data loss. }

As the figure shows, data loss breakdown significantly correlates with both erasure code and SSD type.
The combination of bad disk and bad block (BD+BB) is the dominant source of data loss when using PMDS erasure code. A relatively smaller share of DL, less than 10\% in all SSD types, is caused by co-incidence of two bad blocks (BB+BB) in a data stripe. 
Hence, we can conclude that bad blocks are the dominant source of data loss when using PMDS codes. 
This observation is described by the fact that PMDS codes can correct the combination of bad chip with a bad symbol, and can also correct the coincidence of two bad symbols. 
Bad symbols leading to data loss just happen either in the case two bad symbols in a single stripe coincide with a bad chip (BD+BS), or three bad symbols coincide in a single stripe (BS+BS), that are not so probable (BD+BS and BS+BS failures happened respectively in 123 and two cases of data loss, compared to {12,194,444} cases caused by BD+BB). 
PMDS codes, however, fail to correct the combination of bad disk and bad block, as the bad blocks make an entire data chunk lost, rather than a single symbol (each data chunk includes 4 symbols, considering 4KB page size, 128KB stripe size, and 8 devices per stripe). 
Please note that other combinations of failures also have non-zero values, but are not reported as their contribution {is} less than 1\%.  

Data loss breakdown of RAID5 and RAID6 is more sensitive to SSD type. However, RAID5 failures are dominantly caused by the coincidence of bad chip with either bad block or bad symbol, while bad chips combined with bad blocks cause more than 50\% of data loss in all SSD types. 
In the case of RAID6, the coincidence of bad chip, bad block and bad symbol (BD+BB+BS) has also a significant contribution in total data loss, even greater than BD+BB (caused by one bad chip combined with two bad blocks) for MLCA, MLCC, and MLCD.

\subsection{Impact of Workload}
\begin{figure*}[!htb]
	\captionsetup[subfigure]{labelformat=empty}
	\centering
	\subfloat[]{\includegraphics[width=.5\textwidth]{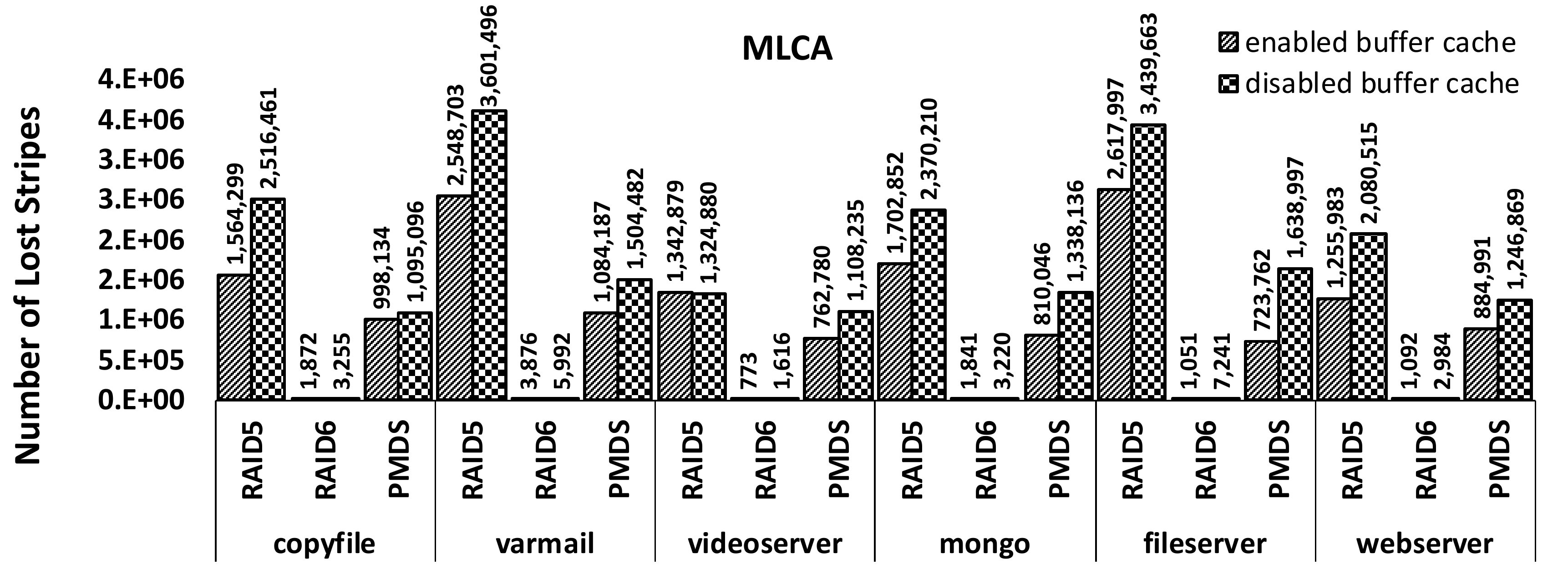}}%
	\hfil
	\subfloat[]{\includegraphics[width=.5\textwidth]{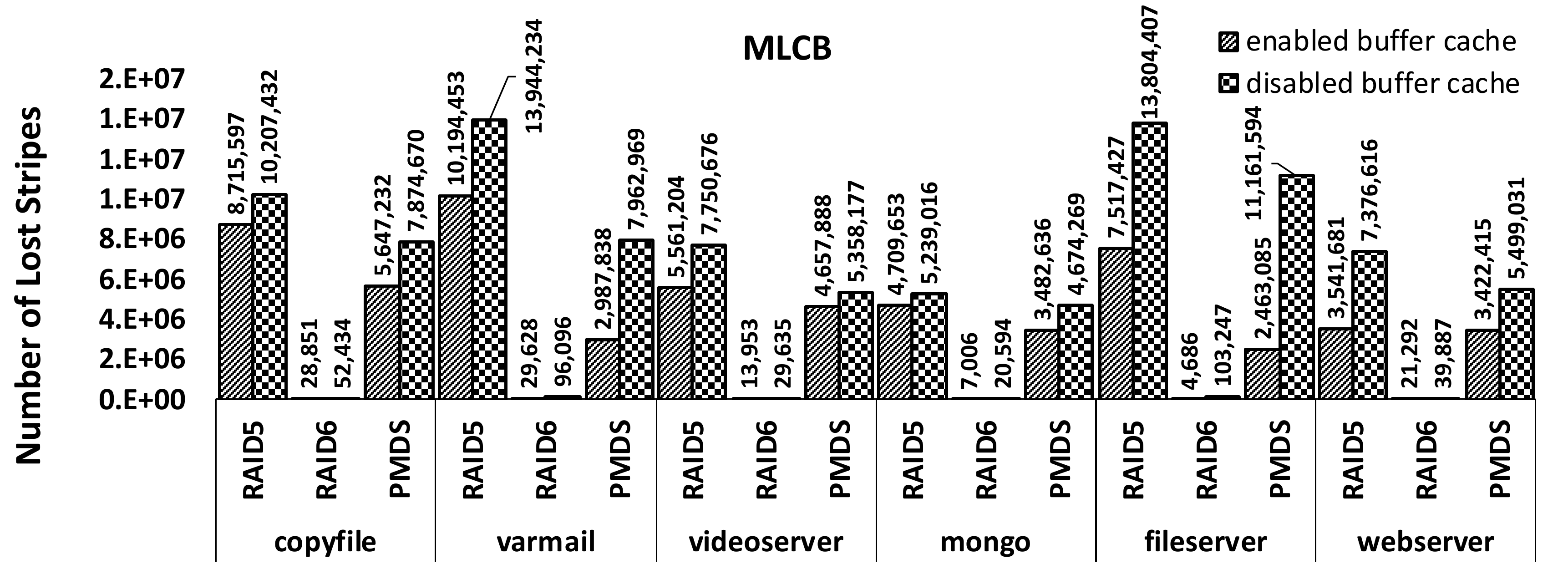}}%
	\hfil
	\vspace{-1.2em}
	\subfloat[]{\includegraphics[width=.5\textwidth]{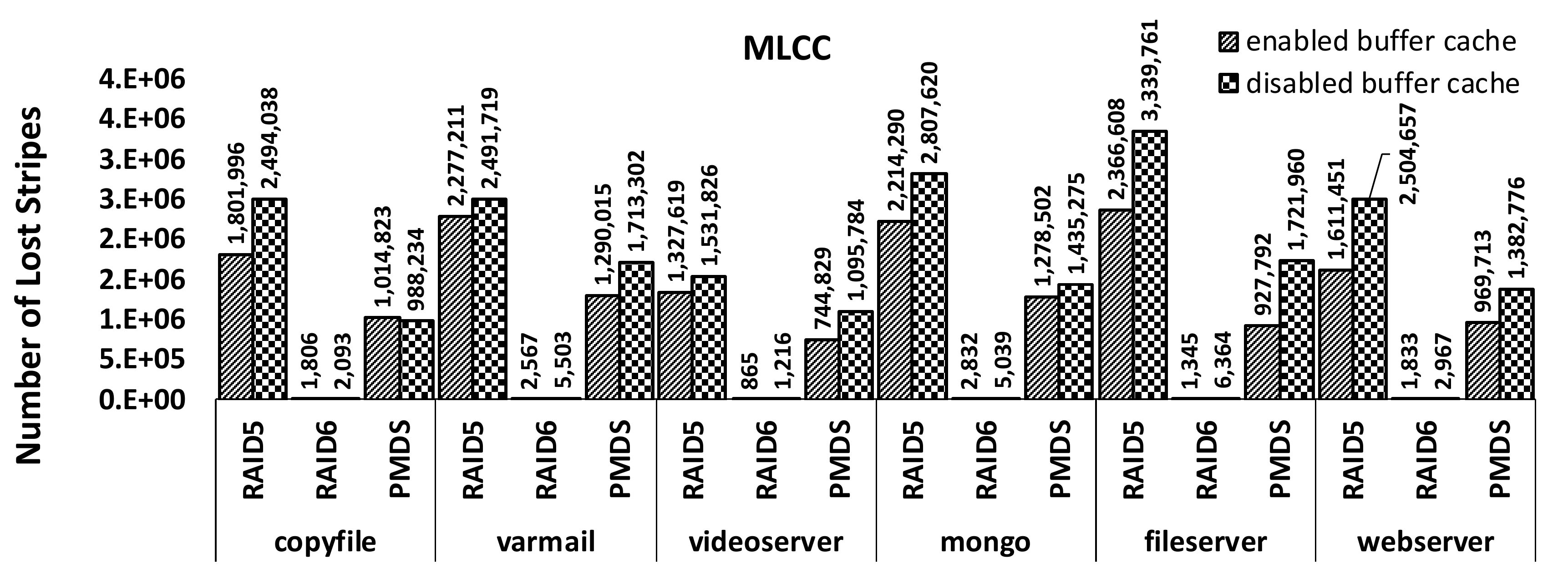}}%
	\hfil
	\subfloat[]{\includegraphics[width=.5\textwidth]{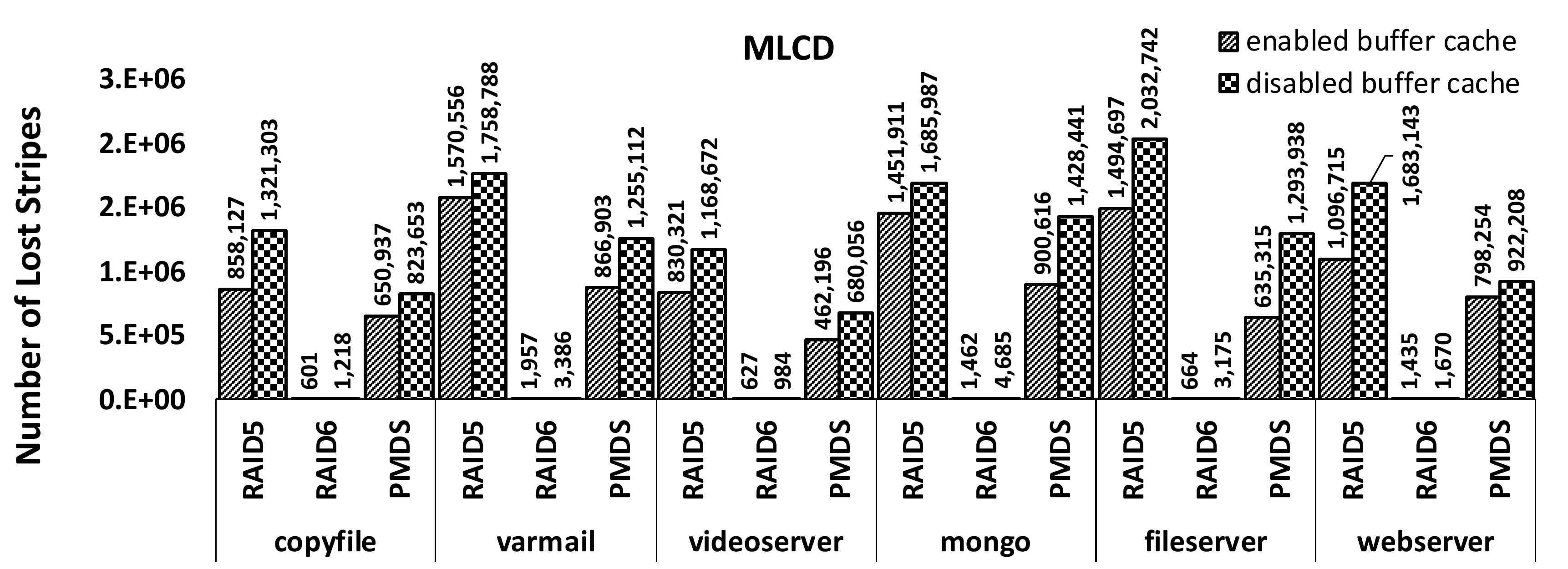}}%
	\hfil
	\vspace{-1.2em}
	\subfloat[]{\includegraphics[width=.5\textwidth]{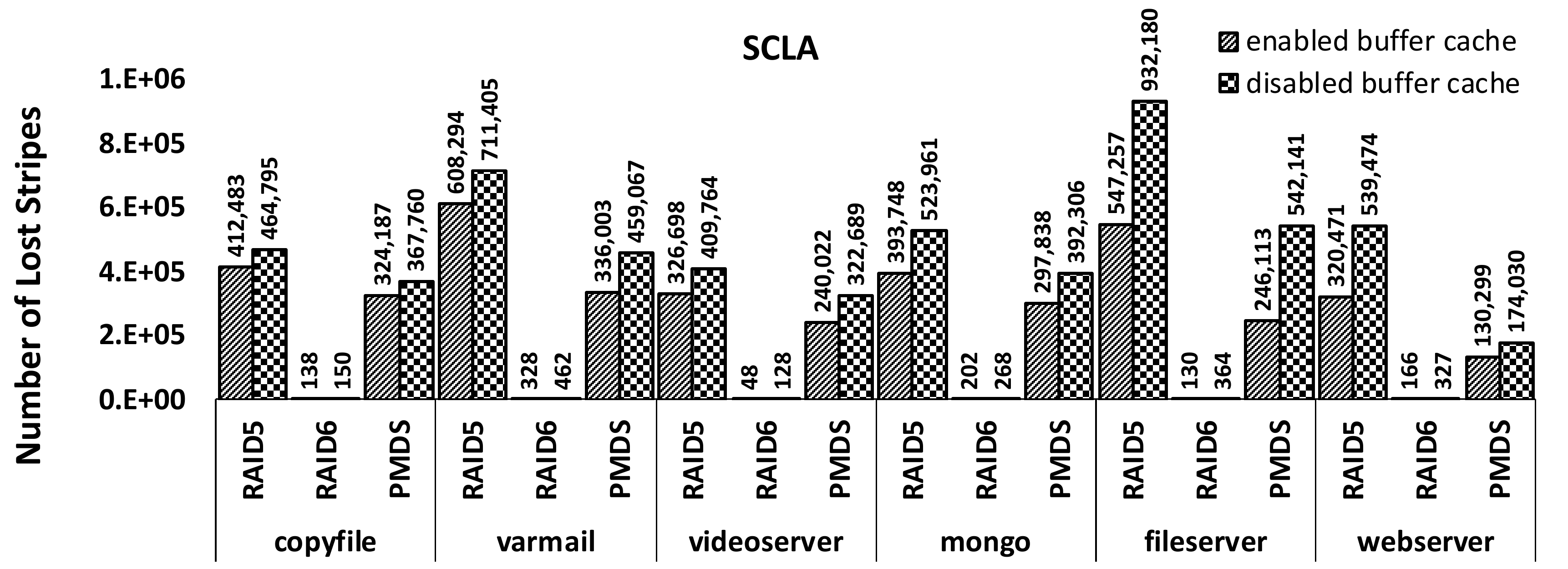}}%
	\hfil
	\subfloat[]{\includegraphics[width=.5\textwidth]{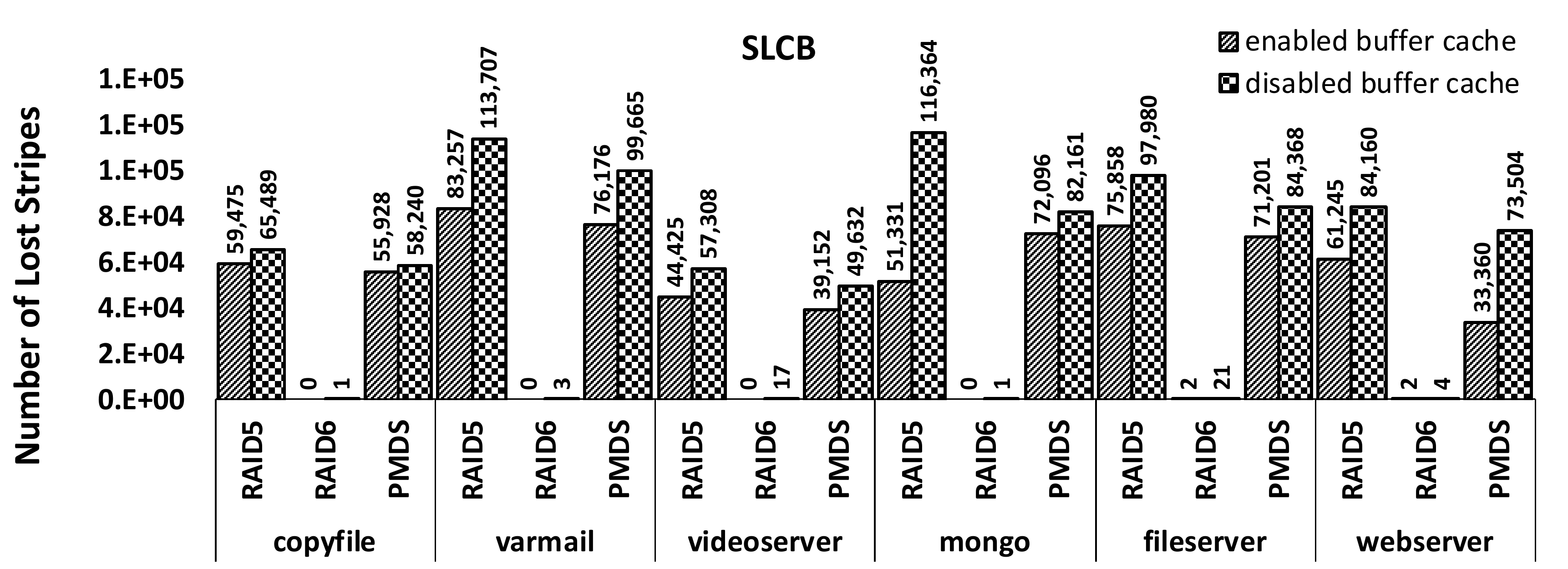}}%
	\vspace{-1.2em}
	\caption{{Comparing reliability of different workloads (TTS = 10,000h, TTR = 10h).}}
	\label{fig:compare-workloads}
\end{figure*}

{Fig.~\ref{fig:compare-workloads} reports the number of lost stripes within 4 years mission time experienced in 1000 SSD arrays in both cases of enabled and disabled buffer cache. The results are reported for different SLC and MLC types. 
SSD arrays experience different magnitude of data loss depending on the examined workload and SSD type. This difference is mainly caused by workload characteristics including number of {P/E} cycles and disk accesses which would be reduced in case of enabling buffer cache.}

An important observation is that the relative reliability of workloads may change in different SSD types. 
For example in MLCA using RAID5 configuration, Fileserver workload experiences the most data loss, while in MLCB the most data loss is  experienced in Varmail workload. 
Different rate of bad chip, bad block, and bad symbol in different SSD types and how they correlate with the workload characteristics is the major source of this observation. 
{While the rate of bad chip and bad block is characterized by SSD type (Table~\ref{tab:field-failure-statistics}) and determined in the start of simulation (discussed in Section~\ref{sec:fault-injection-parameters}), the rate of bad symbol, determined by RBER, is also a function of {P/E} cycles (Fig.~\ref{fig:ssd failure statistics}) and is highly correlated with workload and the impact of buffer cache where by enabling buffer cache we observe about 26.2\%, 56.1\%, and 29.5\% smaller failure rate in RAID5, RAID6, and PMDS, respectively.} 
Accordingly, the workloads characterized by large number of {P/E} cycles ({i.e.,} the workloads dominated by write requests) experience a relatively greater data loss in SSD types with large RBER (MLCA, MLCB, MLCC, and MLCD).
%
%

\subsection{Impact of SSD type}
\begin{figure}
    \centering
 
        \includegraphics[width=0.3\textwidth]{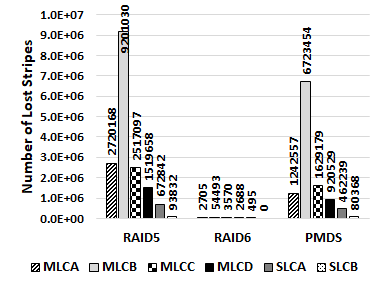}
    \caption{Comparing reliability of different SSD types (TTS = 10,000h, TTR = 10h)}
 \vspace{-0.4cm}
    \label{fig:compare-ssd-types}

\end{figure}

{Fig.~\ref{fig:compare-ssd-types} compares the number of lost stripes in different SSD types. The reported values are aggregated from 11 synthetic and realistic application workloads.} 
As the results show, MLCB is the least reliable SSD, experiencing one order of magnitude greater data loss than SLCB. 
Referring to the failure characteristics of MLCB (Fig.~\ref{fig:ssd failure statistics} and Table~\ref{tab:field-failure-statistics}), this observation is described by MLCB having the greatest RBER, resulting {in} the highest rate of bad symbol between examined SSD types. 
While the mean number of bad blocks (per device) in MLCB is average, it has the greatest percentage of drives with bad blocks (79.5\%), also describing the low reliability of this SSD type. 

Another observation is considerable reliability benefits of SLC types over MLC types, specially in the case of SLCB. 
Both SLCA and SLCB, as Fig.~\ref{fig:ssd failure statistics} shows, have significantly lower RBER compared to MLC types. 
Moreover, percentage of drives with bad chips reported for SLCA and SLCB (Table~\ref{tab:field-failure-statistics}) is considerably lower than MLC types (3.8\% and 2.3\%, respectively for SLCA and SLCB), helping the greater reliability in SLC types. 
While SLCB outperforms SLCA in terms of reliability, Table~\ref{tab:field-failure-statistics} shows that it has greater percentage of drives with bad blocks than SLCA (64.6\% vs 39.0\%). 
This observation is described by the greater percentage of drives with bad chips in SLCA, and slightly greater RBER, compared to SLCB. 

\subsection{Impact of Time to Recover and Time to Scrub}
\begin{figure}
    \centering

        \includegraphics[width=0.48\textwidth]{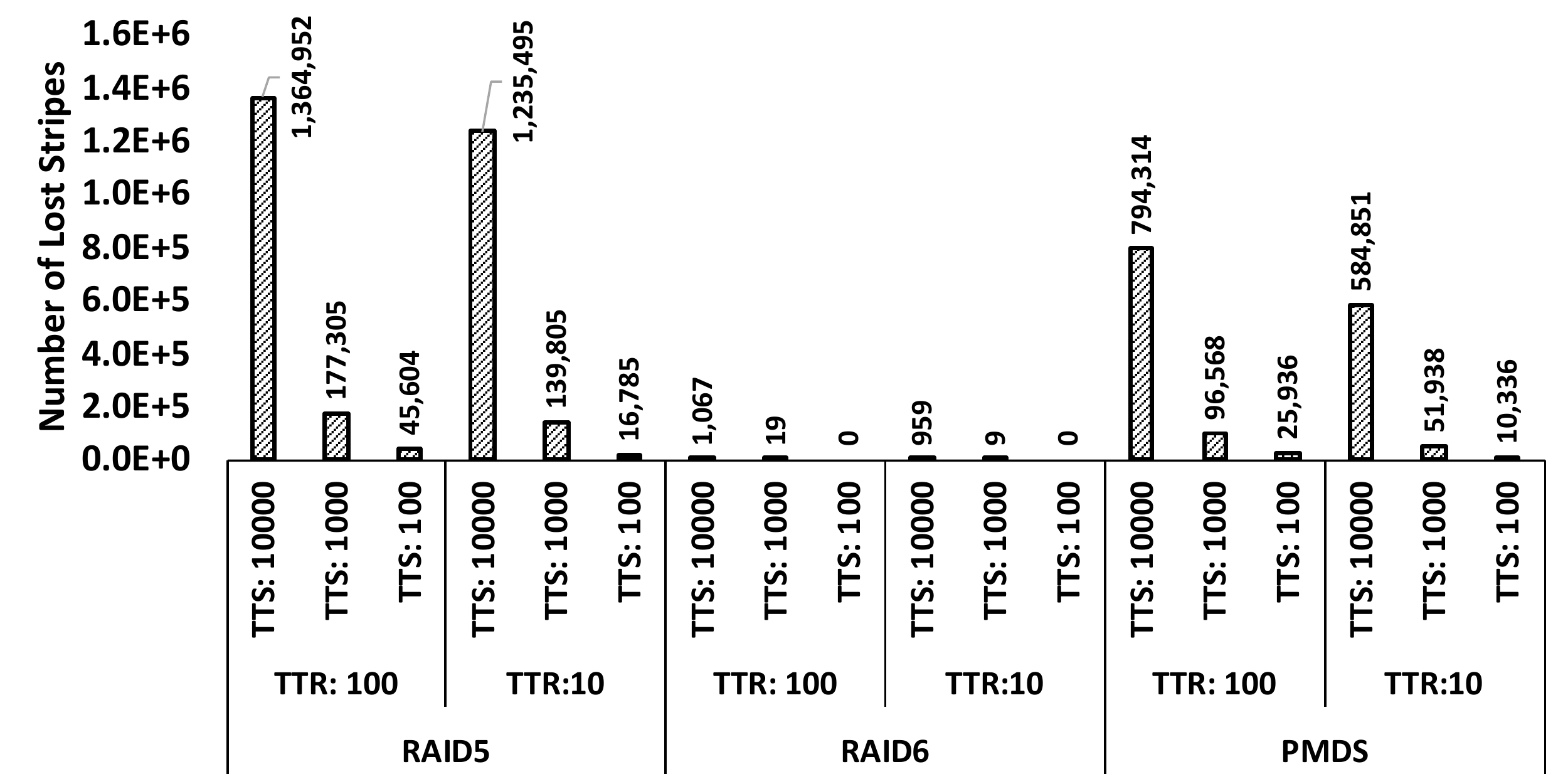}
    \caption{{Impact of TTR (Hours) and TTS (Hours) on the reliability of SSD array (For MLCA under Mixed workload).}}
    \label{fig:ttr-tts-impact}

\end{figure}

Fig.~\ref{fig:ttr-tts-impact} shows the impact of time to recover (TTR) and time to scrub (TTS) on the array reliability. 
One important factor that contributes array reliability is time to recover {the} array from a device failure, by reconstructing the failed device data to a brand-new device. 
Duration of this procedure, however, depends on the array architecture and is a function of parameters such as SSD performance and bandwidth of interconnections. 
Moreover, the {reconstruction} process is usually performed when the array is operational. Hence, the {reconstruction} time is also a function of workload (it takes longer under heavy workloads).   
Increased {reconstruction} time has a negative impact on the array reliability. The reason behind is the accumulation of bit errors within reconstruct process, possibly leading to stripe data loss. 
By increasing TTR from 10 to 100, the expected number of bit error within reconstruct process is increased by 10 times, leading to greater number of lost stripes. 

Another important factor contributing array reliability is time to scrub. 
{Scrubbing} is performed on predefined periods to remove possible bit errors using array redundancies. This process reduces the chance of data loss by preventing the accumulation of bit errors, as well as the combination of device failure and bit errors. 
Scrubbing, however, is a costly process, as it mandates reading and verifying the entire array data. 
Hence, time to scrub is defined to reach {an effective} trade-off between reliability and performance, depending on the policies of datacenter administrators.  

As Fig.~\ref{fig:ttr-tts-impact} shows, the impact of TTS on reliability is significant. 
In the case of RAID5 and PMDS, increasing TTS from 1000 to 10,000 has {an} ascending impact on data loss by almost 8 times. 
This impact is even more drastic in RAID6 and causes 56 times data loss increase. 
In the case of RAID5 and PMDS, increasing TTS from 100 to 10,000 results in 30 times greater data loss. 
It is also worth to mention that the impact of increased TTR is more drastic when having smaller TTS values. 
Under small TTS values, the total number of lost stripes is reduced, magnifying the impact of TTR increase. 
For example in the case of RAID5, increasing TTR from 10 to 100 leads to 10\% data loss increase in the case of TTS=10,000. In the case of TTS=1000 and TTS=100, however, increasing TTR from 10 to 100 results in respectively 26\% and 171\% data loss increase.

{Another important observation is that decreasing TTS improves RAID6 more than RAID5 and PMDS. Better explaining the case, we define \emph{Late Scrub} (LS) a scrubbing process that comes too late to prevent Data Loss (DL) in a stripe and \emph{Late Scrub Threshold} (LST) as the maximum TTS that can prevent DL for each data stripe. We expect weak erasure codes such as RAID5 having lower average LST and powerful erasure codes such as RAID6 having greater average LST. When LST is too low, even a big improvement (decrease) in TTS makes no difference. That is why we here observe decreasing TTS improves RAID6 more than RAID5 and PMDS. }

\subsection{Impact of Stripe Size}
\begin{figure}
    \centering
 
        \includegraphics[width=0.4\textwidth]{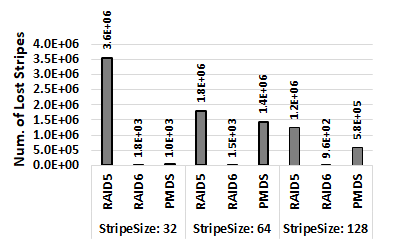}
            \vspace{-0.1cm}
    \caption{{Impact of Stripe Size (32KB, 64KB, and 128KB) on the reliability of SSD array, for MLCA under Mixed workload (TTS=10,000h, TTR=10h). The rest of SSD array parameters are appeared in Table~\ref{tab:ssd array configuration}.} }
    \label{fig:stripe-size-impact}

\end{figure}

Fig.~\ref{fig:stripe-size-impact} shows the impact of stripe size on the array reliability. 
The minimum and maximum possible values for stripe size are determined by RAID controller manufacturers. For SSD RAID controllers, the minimum possible configuration is 64KB~\cite{lsi_raid}. 
However, here we also examine 32KB stripe size configuration in our fault-injection experiments.

Regarding our analysis in Section~\ref{sec:proposedmodel}, the stripe size has no impact on RAID5 and RAID6 codes, as those codes just employ row-wise parity codes. 
PMDS codes, however, benefit smaller stripe size. By reducing the stripe size, the global parity symbol would be responsible for error correction of a smaller number of data symbols, having less chance of fault accumulation leading to uncorrectable error.
The empirical results also confirm our hypothesis and show a significant reliability improvement in PMDS codes when reducing stripe size to 32KB.    
Indeed, when reducing stripe size to 32KB, we {observe} 1002 stripe loss events in PMDS codes, versus 1754 events {that we observe} in RAID6. 
Theoretically, PMDS should not perform better than RAID6 in terms of reliability, but this observation is described by greater write overhead of RAID6 compared to PMDS;  
For the Mixed workload, due to having larger number of writes in RAID6 array (depicted in Fig.~\ref{fig:endurance}), the number of P/E cycles, as well as number of accesses is increased, leading to a greater RBER.
For RAID5 and RAID6, however, reducing stripe size has no impact on reliability. 

Please note that by reducing the stripe size, total number of stripes is doubled. 
Hence, the magnitude of data loss caused by two lost stripes in 32KB mode is equal to the magnitude of data loss caused by one lost stripe in 64KB mode.    
We observe that reducing stripe size by a factor of two almost doubles the number of lost stripes in RAID5 and RAID6, as expected. 
For example, by reducing stripe size from 64KB to 32KB, the number of lost stripes in RAID5 is increased from {$1,810,334$ to $3,562,789$}.  

\subsection{Comparison with Previous Models}
\begin{figure}
    \centering
 
        \includegraphics[width=0.5\textwidth]{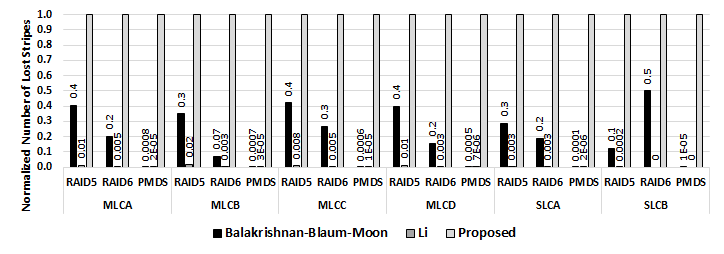}
    \caption{{Comparing previous SSD array reliability models with the proposed model. The chart reports number of lost stripes normalized to proposed model results for TTS=10,000h and TTR=10h. This chart reports the aggregate of copyfile, varmail, videoserver, mango, fileserver, and webserver workloads.}}
 \vspace{-0.2cm}
    \label{fig:previous-models}

\end{figure}

Fig.~\ref{fig:previous-models} compares previous SSD array reliability models with the proposed model. 
{The chart reports number of lost stripes normalized to proposed model results for TTS=10,000h and TTR=10h.} 
In this {chart,} we classify the previous works into two categories.
The models proposed by Balakrishnan et al.~\cite{balakrishnan2010differential}, Blaum et al.~\cite{blaum2013partial}, and Moon et al.~\cite{moon2016does} that consider the coincidence of bad chip and bad symbol and ignore the impact of bad block (as summarized in Table~\ref{tab:qualitative-comparison}) are classified as \emph{Balakrishnan-Blaum-Moon}. 
The model of Li et al.~\cite{li2016analysis} that just takes the coincidence of bad symbols into account (ignores bad chip and bad block) is classified as \emph{Li}. 
The previous works, however, have also other sources of inaccuracy, neglected in this comparison, such as using deprecated SSD failure field data, using either Markov models or closed-form probability equations, and not using real-system implementation.   

As the results show, \emph{Li} provides the less accurate results due to ignoring both bad chips and bad blocks and just considering the coincidence of bad symbols in a data stripe. 
Depending on the SSD type, the results of \emph{Li} underestimate reliability by at least two orders of magnitude.
The results of \emph{Balakrishnan-Blaum-Moon} is more accurate, due to considering the coincidence of bad chips and bad symbols. 
However, in the case of SLCB arrays configured by PMDS codes, we observe \emph{Balakrishnan-Blaum-Moon} underestimates reliability by five orders of magnitude.

\subsection{Summary of Observations}
{Table \ref{table:summary} reports a summary of our observations comparing RAID5, RAID6, and PMDS in terms of source of failures and the impact of array parameters such as SSD type, TTS, TTR, and stripe size.}
\begin{table}[!htb]
	\caption{{Summary of observations comparing RAID5, RAID6, and PMDS.}}
	\vspace{-0.2cm}
	\label{table:summary}
	\begin{adjustbox}{width=0.5\textwidth,totalheight=\textheight,keepaspectratio}
	\begin{tabular}{|l||l|c|c|c|c|}
		\hline
		& \multicolumn{1}{c|}{\begin{tabular}[c]{@{}c@{}}Main Source\\ of Failure\end{tabular}} & \begin{tabular}[c]{@{}c@{}}Dep.\\ to SSD\\ Type\end{tabular} & \begin{tabular}[c]{@{}c@{}}Dep.\\ to TTS\end{tabular} & \begin{tabular}[c]{@{}c@{}}Dep.\\ to TTR\end{tabular} & \begin{tabular}[c]{@{}c@{}}Dep.\\ to Stripe\\Size\end{tabular} \\ \hline\hline
		RAID5 & \begin{tabular}[c]{@{}l@{}}1) BD+BB\\ 2) BD+BS\end{tabular}                           & Yes                                                              & Significant                                                 & \begin{tabular}[c]{@{}c@{}}Less \\ Significant\end{tabular} & No                                                                  \\ \hline
		RAID6 & \begin{tabular}[c]{@{}l@{}}1) BD+BB+BS\\ 2) BD+BB\end{tabular}                        & Yes                                                              & Significant                                                 & \begin{tabular}[c]{@{}c@{}}Less\\ Significant\end{tabular}  & No                                                                  \\ \hline
		PMDS  & BD+BB                                                                                 & Yes                                                              & Significant                                                 & \begin{tabular}[c]{@{}c@{}}Less\\ Significant\end{tabular}  & Yes                                                                 \\ \hline
	\end{tabular}
	\end{adjustbox}
\end{table}

{The detailed observations as reported in Table \ref{table:summary} are as follows:}
\begin{itemize}
\item
Having slightly greater ERF than RAID5, PMDS(1,1) codes are proposed to offer a reliability close to RAID6~\cite{plank2014sector,blaum2013partial} by correcting the combination of device and symbol failures. However, our analysis using recent field results show that PMDS reliability is far behind RAID6. The major source of misleading conclusions in previous works is taking deprecated assumptions about failure characteristics of SSD devices, falsified by state-of-the-art field data~\cite{schroeder2016flash,meza2015large}.
\item
While PMDS(1,1) copes with the combination of device failure and symbol failure, it fails to correct errors combined by device failure and block failure, contributing more than 90\% of total data loss.
\item
Even in RAID5 which can tolerate just a single device failure, the contribution of errors combined by device failure and block failure is more than those combined by device failure and symbol failure. While the rate of block failures is significantly lower than symbol failures, this observation is described by the greater magnitude of data loss imposed by block failures (in our experiments a single SSD block is shared upon 16 data stripes).
\item
The contribution of bad blocks combined by device failure (two bad blocks and one bad device) is also significant in total data loss of RAID6. 
\item
In the resolution of one single data stripe where erasure codes take effect, the block failures manifest as device failure (they result in the loss of a full data chunk, rather than a single symbol). Hence, symbol-level protections (suggested by PMDS codes) are not effective in dealing with block failures. Regarding the significant contribution of bad blocks combined with bad devices in total data loss, we can conclude that device-level protections, such as conventional RAIDs, are the most effective choices at least for the contemporary SSD architectures. 
\item
The dark side of PMDS codes, ignored by its creators~\cite{blaum2013partial,plank2014sector,li2014stair}, is the negative effect of global parity write overhead on SSD endurance, that by itself violates reliability.
\item
SSD array reliability, as well as the failure breakdown, is significantly correlated with SSD type. 
\item
Previous models on the reliability of SSD arrays just focus on the coincidence of bad symbols and bad chip. Our study, however, shows that this type of failure contributes the minority of data loss in SSD arrays and the previous models underestimate data loss by less than half.
\item
Time to scrub has a significant impact on array reliability, while the impact of time to recover from a device failure is of less significance. 
\item
RAID5 and RAID6 codes which use row-wise parity, perform almost independent of stripe size. PMDS codes, however, benefit smaller stripe sizes and show a promising reliability improvement when reducing stripe size from 128KB to 32KB.
This observation motivates us for further investigations of the effect of stripe size, under different workloads and array architectures. 
\end{itemize}

\section{Conclusion}
\label{sec:conclusion}
{In this paper, we investigated the reliability of SSD arrays using real-system implementation of conventional and emerging erasure codes, under realistic storage traces.}
The reliability is evaluated by statistical fault injection experiments that {post-process} the SSD usage logs obtained from the real-system implementation, 
while the fault/failure attributes are obtained from the state-of-the-art field data by previous works. 
As a case study, we examined conventional RAID5 and RAID6 and emerging PMDS codes, SD codes and STAIR codes in terms of both reliability and performance using an open-source software RAID controller, MD (in Linux kernel version 3.10.0-327), and arrays of Samsung 850 Pro SSDs. 

Our experiments showed that previous models underestimate the SSD array reliability by up to six orders of magnitude, as they just focus on the coincidence of bad pages (bit errors) and bad chips within a data stripe that holds the minority of data loss cause in SSD arrays. 
We observed the combination of bad chips with bad blocks as the major source of data loss in RAID5 and emerging codes (contributing more than 54\% and 90\% of data loss in RAID5 and emerging codes, respectively), while RAID6 remained robust under these failure combinations.
We also observed that time to scrub is a significant contributor to array data loss, while the impact of time to recover from a device failure is of less significance. 
Finally, the fault injection results show that SSD array reliability, as well as the failure breakdown, is significantly correlated with SSD type.
While our empirical results showed that emerging erasure codes fail to replace RAID6 in terms of reliability when having stripe sizes commonly used in enterprise RAID controllers (128KB and 64KB), for a speculative 32KB stripe size we observed a promising reliability improvement in emerging erasure codes, performing similar to RAID6.





\bibliographystyle{IEEEtran}
\bibliography{comparing_erasure_codes_13_12_2019_main}

\begin{thebibliography}{10}
\providecommand{\url}[1]{#1}
\csname url@samestyle\endcsname
\providecommand{\newblock}{\relax}
\providecommand{\bibinfo}[2]{#2}
\providecommand{\BIBentrySTDinterwordspacing}{\spaceskip=0pt\relax}
\providecommand{\BIBentryALTinterwordstretchfactor}{4}
\providecommand{\BIBentryALTinterwordspacing}{\spaceskip=\fontdimen2\font plus
\BIBentryALTinterwordstretchfactor\fontdimen3\font minus
  \fontdimen4\font\relax}
\providecommand{\BIBforeignlanguage}[2]{{%
\expandafter\ifx\csname l@#1\endcsname\relax
\typeout{** WARNING: IEEEtran.bst: No hyphenation pattern has been}%
\typeout{** loaded for the language `#1'. Using the pattern for}%
\typeout{** the default language instead.}%
\else
\language=\csname l@#1\endcsname
\fi
#2}}
\providecommand{\BIBdecl}{\relax}
\BIBdecl

\bibitem{demara2018non}
R.~F. DeMara and P.~Montuschi, ``Non-volatile memory trends: Toward improving
  density and energy profiles across the system stack,'' \emph{Computer},
  no.~4, pp. 12--13, 2018.

\bibitem{wang2018rc}
P.~Wang, S.~Li, G.~Sun, X.~Wang, Y.~Chen, H.~Li, J.~Cong, N.~Xiao, and
  T.~Zhang, ``Rc-nvm: Enabling symmetric row and column memory accesses for
  in-memory databases,'' in \emph{High Performance Computer Architecture
  (HPCA)}.\hskip 1em plus 0.5em minus 0.4em\relax Vienna, Austria: IEEE, 2018,
  pp. 518--530.

\bibitem{henkel2017emerging}
J.~Henkel, ``Emerging memory technologies,'' \emph{IEEE Design \& Test},
  vol.~34, no.~3, pp. 4--5, 2017.

\bibitem{huai2008spin}
Y.~Huai, ``Spin-transfer torque mram (stt-mram): Challenges and prospects,''
  \emph{Association of Asia Pacific Physical Societies (AAPPS) bulletin},
  vol.~18, no.~6, pp. 33--40, 2008.

\bibitem{yang2014improving}
C.~Yang, H.-M. Chen, T.~N. Mudge, and C.~Chakrabarti, ``Improving the
  reliability of mlc nand flash memories through adaptive data refresh and
  error control coding,'' \emph{Journal of Signal Processing Systems}, vol.~76,
  no.~3, pp. 225--234, 2014.

\bibitem{yang2011flexible}
C.~Yang, Y.~Emre, C.~Chakrabarti, and T.~Mudge, ``Flexible product code-based
  ecc schemes for mlc nand flash memories,'' in \emph{Signal Processing Systems
  (SiPS)}.\hskip 1em plus 0.5em minus 0.4em\relax Beirut, Lebanon: IEEE, 2011,
  pp. 255--260.

\bibitem{khoshavi2018read}
N.~Khoshavi and R.~F. DeMara, ``Read-tuned stt-ram and edram cache hierarchies
  for throughput and energy optimization,'' \emph{IEEE Access}, vol.~6, pp.
  14\,576--14\,590, 2018.

\bibitem{fedorov2017speculative}
V.~Fedorov, J.~Kim, M.~Qin, P.~V. Gratz, and A.~Reddy, ``Speculative paging for
  future nvm storage,'' in \emph{International Symposium on Memory Systems
  (MEMSYS)}.\hskip 1em plus 0.5em minus 0.4em\relax Alexandria, VA, USA: ACM,
  2017, pp. 399--410.

\bibitem{qin2011mnftl}
Z.~Qin, Y.~Wang, D.~Liu, Z.~Shao, and Y.~Guan, ``Mnftl: An efficient flash
  translation layer for mlc nand flash memory storage systems,'' in
  \emph{Design Automation Conference (DAC)}.\hskip 1em plus 0.5em minus
  0.4em\relax San Diego, CA, USA: ACM, 2011, pp. 17--22.

\bibitem{liu2011pcm}
D.~Liu, T.~Wang, Y.~Wang, Z.~Qin, and Z.~Shao, ``Pcm-ftl: A
  write-activity-aware nand flash memory management scheme for pcm-based
  embedded systems,'' in \emph{Real-Time Systems Symposium (RTSS)}.\hskip 1em
  plus 0.5em minus 0.4em\relax Vienna, Austria: IEEE, 2011, pp. 357--366.

\bibitem{qin2010demand}
Z.~Qin, Y.~Wang, D.~Liu, and Z.~Shao, ``Demand-based block-level address
  mapping in large-scale nand flash storage systems,'' in \emph{IEEE/ACM/IFIP
  International Conference on Hardware/Software Codesign and System Synthesis
  (CODES+ISSS)}.\hskip 1em plus 0.5em minus 0.4em\relax Scottsdale, AZ, USA:
  ACM, 2010, pp. 173--182.

\bibitem{qin2011two}
------, ``A two-level caching mechanism for demand-based page-level address
  mapping in nand flash memory storage systems,'' in \emph{Real-Time and
  Embedded Technology and Applications Symposium (RTAS)}.\hskip 1em plus 0.5em
  minus 0.4em\relax Chicago, IL, USA: IEEE, 2011, pp. 157--166.

\bibitem{dong2014nvsim}
X.~Dong, C.~Xu, N.~Jouppi, and Y.~Xie, ``Nvsim: A circuit-level performance,
  energy, and area model for emerging non-volatile memory,'' \emph{Emerging
  Memory Technologies}, pp. 15--50, 2014.

\bibitem{wu2009hybrid}
X.~Wu, J.~Li, L.~Zhang, E.~Speight, R.~Rajamony, and Y.~Xie, ``Hybrid cache
  architecture with disparate memory technologies,'' \emph{ACM SIGARCH computer
  architecture news}, vol.~37, no.~3, pp. 34--45, 2009.

\bibitem{joo2010energy}
Y.~Joo, D.~Niu, X.~Dong, G.~Sun, N.~Chang, and Y.~Xie, ``Energy-and
  endurance-aware design of phase change memory caches,'' in \emph{Design,
  Automation \& Test in Europe (DATE)}.\hskip 1em plus 0.5em minus 0.4em\relax
  Dresden, Germany: IEEE, 2010, pp. 136--141.

\bibitem{zand2017energy}
R.~Zand, A.~Roohi, and R.~F. DeMara, ``Energy-efficient and
  process-variation-resilient write circuit schemes for spin hall effect mram
  device,'' \emph{IEEE Transactions on Very Large Scale Integration (VLSI)
  Systems}, vol.~25, no.~9, pp. 2394--2401, 2017.

\bibitem{guan2017block}
Y.~Guan, G.~Wang, C.~Ma, R.~Chen, Y.~Wang, and Z.~Shao, ``A block-level
  log-block management scheme for mlc nand flash memory storage systems,''
  \emph{IEEE Transactions on Computers}, vol.~66, no.~9, pp. 1464--1477, 2017.

\bibitem{kang2018reliability}
Y.~Kang, X.~Zhang, Z.~Shao, R.~Chen, and Y.~Wang, ``A reliability enhanced
  video storage architecture in hybrid slc/mlc nand flash memory,''
  \emph{Journal of Systems Architecture}, vol.~88, pp. 33--42, 2018.

\bibitem{wang2015propram}
Y.~Wang, Y.~Han, L.~Zhang, H.~Li, and X.~Li, ``Propram: exploiting the
  transparent logic resources in non-volatile memory for near data computing,''
  in \emph{Design Automation Conference (DAC)}.\hskip 1em plus 0.5em minus
  0.4em\relax San Francisco, CA, USA: ACM, 2015, p.~47.

\bibitem{zhao2016optical}
Q.~Zhao, M.~Rajaei, I.~Krivorotov, M.~Nilsson, N.~Bagherzadeh, and O.~Boyraz,
  ``Optical investigation of radiation induced conductivity changes in stt-ram
  cells,'' in \emph{Conference on Lasers and Electro-Optics (CLEO)}.\hskip 1em
  plus 0.5em minus 0.4em\relax San Jose, CA, USA: IEEE, 2016, pp. 1--2.

\bibitem{salkhordeh2019analytical}
R.~Salkhordeh, O.~Mutlu, and H.~Asadi, ``An analytical model for performance
  and lifetime estimation of hybrid dram-nvm main memories,'' \emph{IEEE
  Transactions on Computers}, vol.~68, no.~8, pp. 1114--1130, 2019.

\bibitem{li2014nitro}
C.~Li, P.~Shilane, F.~Douglis, H.~Shim, S.~Smaldone, and G.~Wallace, ``Nitro: A
  capacity-optimized ssd cache for primary storage.'' in \emph{USENIX Annual
  Technical Conference}, Philadelphia, PA, USA, 2014, pp. 501--512.

\bibitem{li2017pannier}
C.~Li, P.~Shilane, F.~Douglis, and G.~Wallace, ``Pannier: Design and analysis
  of a container-based flash cache for compound objects,'' \emph{ACM
  Transactions on Storage (TOS)}, vol.~13, no.~3, p.~24, 2017.

\bibitem{liu2013duracache}
R.-S. Liu, C.-L. Yang, C.-H. Li, and G.-Y. Chen, ``Duracache: A durable ssd
  cache using mlc nand flash,'' in \emph{Design Automation Conference
  (DAC)}.\hskip 1em plus 0.5em minus 0.4em\relax Austin, TX, USA: ACM, 2013, p.
  166.

\bibitem{tarihi2016hybrid}
M.~Tarihi, H.~Asadi, A.~Haghdoost, M.~Arjomand, and H.~Sarbazi-Azad, ``A hybrid
  non-volatile cache design for solid-state drives using comprehensive i/o
  characterization,'' \emph{IEEE Transactions on Computers}, vol.~65, no.~6,
  pp. 1678--1691, 2016.

\bibitem{reca}
R.~Salkhordeh, S.~Ebrahimi, and H.~Asadi, ``Reca: an efficient reconfigurable
  cache architecture for storage systems with online workload
  characterization,'' \emph{IEEE Transactions on Parallel and Distributed
  Systems (TPDS)}, vol.~PP, no.~PP, pp. 1--1, 2018.

\bibitem{ahmadian2018eci}
S.~Ahmadian, O.~Mutlu, and H.~Asadi, ``Eci-cache: A high-endurance and
  cost-efficient i/o caching scheme for virtualized platforms,''
  \emph{Proceedings of the ACM on Measurement and Analysis of Computing Systems
  (POMACS)}, vol.~2, no.~1, p.~9, 2018.

\bibitem{salkhordeh2018efficient}
R.~Salkhordeh, M.~Hadizadeh, and H.~Asadi, ``An efficient hybrid i/o caching
  architecture using heterogeneous ssds,'' \emph{IEEE Transactions on Parallel
  and Distributed Systems}, vol.~30, no.~6, pp. 1238--1250, 2018.

\bibitem{hu2011performance}
Y.~Hu, H.~Jiang, D.~Feng, L.~Tian, H.~Luo, and S.~Zhang, ``Performance impact
  and interplay of ssd parallelism through advanced commands, allocation
  strategy and data granularity,'' in \emph{International Conference on
  Supercomputing (ICS)}.\hskip 1em plus 0.5em minus 0.4em\relax Tucson,
  Arizona, USA: ACM, 2011, pp. 96--107.

\bibitem{guerra2011cost}
J.~Guerra, H.~Pucha, J.~S. Glider, W.~Belluomini, and R.~Rangaswami, ``Cost
  effective storage using extent based dynamic tiering.'' in \emph{Conference
  on File and Storage Technologies (FAST)}, vol.~11, San Jose, CA, USA, 2011,
  pp. 20--20.

\bibitem{salkhordeh2015operating}
R.~Salkhordeh, H.~Asadi, and S.~Ebrahimi, ``Operating system level data tiering
  using online workload characterization,'' \emph{{The Journal of
  Supercomputing}}, vol.~71, no.~4, pp. 1534--1562, 2015.

\bibitem{elerath2014beyond}
J.~G. Elerath and J.~Schindler, ``Beyond mttdl: A closed-form raid 6
  reliability equation,'' \emph{ACM Transactions on Storage (TOS)}, vol.~10,
  no.~2, p.~7, 2014.

\bibitem{park2009reliability}
K.~Park, D.-H. Lee, Y.~Woo, G.~Lee, J.-H. Lee, and D.-H. Kim, ``Reliability and
  performance enhancement technique for ssd array storage system using raid
  mechanism,'' in \emph{International Symposium on Communications and
  Information Technology (ISCIT)}.\hskip 1em plus 0.5em minus 0.4em\relax
  Incheon, Korea: IEEE, 2009, pp. 140--145.

\bibitem{kishani2019dependability}
M.~Kishani, M.~Tahoori, and H.~Asadi, ``Dependability analysis of data storage
  systems in presence of soft errors,'' \emph{IEEE Transactions on
  Reliability}, vol.~68, no.~1, pp. 201--215, 2019.

\bibitem{kishani-tr-2018}
M.~Kishani and H.~Asadi, ``Modeling impact of human errors on the data
  unavailability and data loss of storage systems,'' \emph{IEEE Transactions on
  Reliability (TR)}, vol.~67, no.~3, pp. 1111--1127, 2018.

\bibitem{patterson1988case}
D.~A. Patterson, G.~Gibson, and R.~H. Katz, ``A case for redundant arrays of
  inexpensive disks (raid),'' vol.~17, no.~3, 1988.

\bibitem{schroeder2016flash}
B.~Schroeder, R.~Lagisetty, and A.~Merchant, ``Flash reliability in production:
  The expected and the unexpected.'' in \emph{Conference on File and Storage
  Technologies (FAST)}.\hskip 1em plus 0.5em minus 0.4em\relax Santa Clara, CA,
  USA: USENIX, 2016, pp. 67--80.

\bibitem{meza2015large}
J.~Meza, Q.~Wu, S.~Kumar, and O.~Mutlu, ``A large-scale study of flash memory
  failures in the field,'' in \emph{ACM SIGMETRICS Performance Evaluation
  Review}, vol.~43, no.~1.\hskip 1em plus 0.5em minus 0.4em\relax Portland,
  Oregon, USA: ACM, 2015, pp. 177--190.

\bibitem{narayanan2016ssd}
I.~Narayanan, D.~Wang, M.~Jeon, B.~Sharma, L.~Caulfield, A.~Sivasubramaniam,
  B.~Cutler, J.~Liu, B.~Khessib, and K.~Vaid, ``Ssd failures in datacenters:
  What? when? and why?'' in \emph{International Systems and Storage
  Conference}.\hskip 1em plus 0.5em minus 0.4em\relax Haifa, Israel: ACM, 2016,
  p.~7.

\bibitem{ahmadian-ssd-rel-date}
S.~Ahmadian, F.~Taheri, M.~Lotfi, M.~Karimi, and H.~Asadi, ``Investigating
  power outage effects on reliability of solid-state drives,'' in \emph{Design,
  Automation and Test in Europe (DATE)}.\hskip 1em plus 0.5em minus 0.4em\relax
  Dresden, Germany: IEEE/ACM, March 2018.

\bibitem{li2016analysis}
Y.~Li, P.~P. Lee, and J.~C. Lui, ``Analysis of reliability dynamics of ssd
  raid,'' \emph{IEEE Transactions on Computers}, vol.~65, no.~4, pp.
  1131--1144, 2016.

\bibitem{kim2013improving}
J.~Kim, J.~Lee, J.~Choi, D.~Lee, and S.~H. Noh, ``Improving ssd reliability
  with raid via elastic striping and anywhere parity,'' in \emph{Dependable
  Systems and Networks (DSN)}.\hskip 1em plus 0.5em minus 0.4em\relax Budapest,
  Hungary: IEEE, 2013, pp. 1--12.

\bibitem{moon2016does}
S.~Moon and A.~Reddy, ``Does raid improve lifetime of ssd arrays?'' \emph{ACM
  Transactions on Storage (TOS)}, vol.~12, no.~3, p.~11, 2016.

\bibitem{balakrishnan2010differential}
M.~Balakrishnan, A.~Kadav, V.~Prabhakaran, and D.~Malkhi, ``Differential raid:
  Rethinking raid for ssd reliability,'' \emph{ACM Transactions on Storage
  (TOS)}, vol.~6, no.~2, p.~4, 2010.

\bibitem{greenan2009building}
K.~M. Greenan, D.~D. Long, E.~L. Miller, T.~Schwarz, and A.~Wildani, ``Building
  flexible, fault-tolerant flash-based storage systems,'' in \emph{Proceedings
  of the 5th Workshop on Hot Topics in System Dependability (HotDep 2009)},
  Yokohama, Japan, 2009.

\bibitem{blaum2013partial}
M.~Blaum, J.~L. Hafner, and S.~Hetzler, ``Partial-mds codes and their
  application to raid type of architectures,'' \emph{IEEE Transactions on
  Information Theory}, vol.~59, no.~7, pp. 4510--4519, 2013.

\bibitem{grupp2012bleak}
L.~M. Grupp, J.~D. Davis, and S.~Swanson, ``The bleak future of nand flash
  memory,'' in \emph{Conference on File and Storage Technologies (FAST)}.\hskip
  1em plus 0.5em minus 0.4em\relax San Jose, CA, USA: USENIX, 2012, pp. 2--2.

\bibitem{cai2012error}
Y.~Cai, E.~F. Haratsch, O.~Mutlu, and K.~Mai, ``Error patterns in mlc nand
  flash memory: Measurement, characterization, and analysis,'' in \emph{Design,
  Automation and Test in Europe}.\hskip 1em plus 0.5em minus 0.4em\relax
  Dresden, Germany: EDA Consortium, 2012, pp. 521--526.

\bibitem{grupp2009characterizing}
L.~M. Grupp, A.~M. Caulfield, J.~Coburn, S.~Swanson, E.~Yaakobi, P.~H. Siegel,
  and J.~K. Wolf, ``Characterizing flash memory: anomalies, observations, and
  applications,'' in \emph{Microarchitecture, Annual IEEE/ACM International
  Symposium on (MICRO)}.\hskip 1em plus 0.5em minus 0.4em\relax New York, NY,
  USA: IEEE, 2009, pp. 24--33.

\bibitem{plank2014sector}
J.~S. Plank and M.~Blaum, ``Sector-disk (sd) erasure codes for mixed failure
  modes in raid systems,'' \emph{ACM Transactions on Storage (TOS)}, vol.~10,
  no.~1, p.~4, 2014.

\bibitem{li2014stair}
M.~Li and P.~P. Lee, ``Stair codes: A general family of erasure codes for
  tolerating device and sector failures,'' \emph{ACM Transactions on Storage
  (TOS)}, vol.~10, no.~4, p.~14, 2014.

\bibitem{mielke2008bit}
N.~Mielke, T.~Marquart, N.~Wu, J.~Kessenich, H.~Belgal, E.~Schares, F.~Trivedi,
  E.~Goodness, and L.~R. Nevill, ``Bit error rate in nand flash memories,'' in
  \emph{IEEE International Reliability Physics Symposium (IRPS)}.\hskip 1em
  plus 0.5em minus 0.4em\relax Phoenix, AZ, USA: IEEE, 2008, pp. 9--19.

\bibitem{greenan2007disaster}
K.~M. Greenan, E.~L. Miller, T.~J. Schwarz, and D.~D. Long, ``Disaster recovery
  codes: increasing reliability with large-stripe erasure correcting codes,''
  in \emph{ACM workshop on Storage security and survivability}.\hskip 1em plus
  0.5em minus 0.4em\relax Alexandria, VA, USA: ACM, 2007, pp. 31--36.

\bibitem{paris2007self}
J.-F. P{\^a}ris, T.~J. Schwarz, and D.~D. Long, ``Self-adaptive two-dimensional
  raid arrays,'' in \emph{International Performance, Computing, and
  Communications Conference (IPCCC)}.\hskip 1em plus 0.5em minus 0.4em\relax
  New Orleans, Louisiana, USA: IEEE, 2007, pp. 246--253.

\bibitem{rincon2017disk}
C.~C.~A. Rinc{\'o}n, J.-F. P{\^a}ris, R.~Vilalta, A.~M. Cheng, and D.~D. Long,
  ``Disk failure prediction in heterogeneous environments,'' in \emph{Symposium
  on Performance Evaluation of Computer and Telecommunication Systems
  (SPECTS)}.\hskip 1em plus 0.5em minus 0.4em\relax Seattle, WA, USA: IEEE,
  2017, pp. 1--7.

\bibitem{schwarz2016resar}
T.~Schwarz, A.~Amer, T.~Kroeger, E.~Miller, D.~Long, and J.-F. P{\^a}ris,
  ``Resar: Reliable storage at exabyte scale,'' in \emph{Modeling, Analysis and
  Simulation of Computer and Telecommunication Systems (MASCOTS)}.\hskip 1em
  plus 0.5em minus 0.4em\relax London, UK: IEEE, 2016, pp. 211--220.

\bibitem{paris2013three}
J.-F. P{\^a}ris, D.~D. Long, and W.~Litwin, ``Three-dimensional redundancy
  codes for archival storage,'' in \emph{Modeling, Analysis \& Simulation of
  Computer and Telecommunication Systems (MASCOTS)}.\hskip 1em plus 0.5em minus
  0.4em\relax San Francisco, CA, USA: IEEE, 2013, pp. 328--332.

\bibitem{kao2013flexible}
H.-W. Kao, J.-F. P{\^a}ris, S.~T. Schwarz, and D.~D. Long, ``A flexible
  simulation tool for estimating data loss risks in storage arrays,'' in
  \emph{Mass Storage Systems and Technologies (MSST)}.\hskip 1em plus 0.5em
  minus 0.4em\relax Long Beach, CA, USA: IEEE, 2013, pp. 1--5.

\bibitem{gopinath2009reliability}
K.~Gopinath, J.~Elerath, and D.~Long, ``Reliability modelling of disk
  subsystems with probabilistic model checking,'' University of California,
  Santa Cruz, Tech. Rep., 2009.

\bibitem{paris2008mttdls}
J.-F. P{\^a}ris, T.~J. Schwarz, D.~D. Long, and A.~Amer, ``When mttdls are not
  good enough: Providing better estimates of disk array reliability,'' in
  \emph{International Information and Telecommunication Technologies
  Symposium}, Foz do Iguaçú, Brazil, 2008, pp. 140--145.

\bibitem{kishani2017}
M.~Kishani, R.~Eftekhari, and H.~Asadi, ``Evaluating impact of human errors on
  the availability of data storage systems,'' in \emph{Design, Automation and
  Test in Europe Conference (DATE)}.\hskip 1em plus 0.5em minus 0.4em\relax
  Lausanne, Switzerland: IEEE/ACM, 2017.

\bibitem{e2000transient}
E.~d.~S. e~Silva and H.~R. Gail, ``Transient solutions for markov chains,'' in
  \emph{Computational Probability}.\hskip 1em plus 0.5em minus 0.4em\relax
  Springer, 2000, pp. 43--79.

\bibitem{agrawal2008design}
N.~Agrawal, V.~Prabhakaran, T.~Wobber, J.~D. Davis, M.~S. Manasse, and
  R.~Panigrahy, ``Design tradeoffs for ssd performance.'' in \emph{USENIX
  Annual Technical Conference}, Boston, Massachusetts, USA, 2008, pp. 57--70.

\bibitem{bucy2008disksim}
J.~S. Bucy, J.~Schindler, S.~W. Schlosser, and G.~R. Ganger, ``The disksim
  simulation environment version 4.0 reference manual (cmu-pdl-08-101),''
  \emph{Parallel Data Laboratory}, p.~26, 2008.

\bibitem{kim4ds}
J.~Kim, J.~Lee, J.~Choi, D.~Lee, and S.~H. Noh, ``Ds-raid: Efficient parity
  update scheme for ssds,'' in \emph{Conference on File and Storage
  Technologies (FAST)}, vol.~4.\hskip 1em plus 0.5em minus 0.4em\relax San
  Jose, CA: USENIX, 2012, p.~2.

\bibitem{wicker1995error}
S.~B. Wicker, \emph{Error control systems for digital communication and
  storage}.\hskip 1em plus 0.5em minus 0.4em\relax Prentice hall Englewood
  Cliffs, 1995, vol.~1.

\bibitem{kishani2011hvd}
M.~Kishani, H.~R. Zarandi, H.~Pedram, A.~Tajary, M.~Raji, and B.~Ghavami,
  ``Hvd: horizontal-vertical-diagonal error detecting and correcting code to
  protect against with soft errors,'' \emph{Design Automation for Embedded
  Systems}, vol.~15, no. 3-4, pp. 289--310, 2011.

\bibitem{kishani2011using}
M.~Kishani, A.~Baniasadi, and H.~Pedram, ``Using silent writes in low-power
  traffic-aware ecc,'' in \emph{International Workshop on Power and Timing
  Modeling, Optimization and Simulation (PATMOS)}.\hskip 1em plus 0.5em minus
  0.4em\relax Madrid, Spain: Springer, 2011, pp. 180--192.

\bibitem{schwarz2004disk}
T.~J. Schwarz, Q.~Xin, E.~L. Miller, D.~D. Long, A.~Hospodor, and S.~Ng, ``Disk
  scrubbing in large archival storage systems,'' in \emph{Modeling, Analysis,
  and Simulation of Computer and Telecommunications Systems (MASCOTS)}.\hskip
  1em plus 0.5em minus 0.4em\relax Volendam, Netherlands: IEEE, 2004, pp.
  409--418.

\bibitem{rothberg2005disk}
M.~S. Rothberg, ``Disk drive for receiving setup data in a self monitoring
  analysis and reporting technology (smart) command,'' 2005, uS Patent
  6,895,500.

\bibitem{blktrace}
\BIBentryALTinterwordspacing
(2017) {blktrace: A Block Layer IO Tracing Tool}. [Online]. Available:
  \url{https://www.cse.unsw.edu.au/~aaronc/iosched/doc/blktrace.html}
\BIBentrySTDinterwordspacing

\bibitem{fio}
\BIBentryALTinterwordspacing
(2017) {Fio: Flexible I/O Tester Synthetic Benchmark}. [Online]. Available:
  \url{https://github.com/axboe/fio}
\BIBentrySTDinterwordspacing

\bibitem{tarasov2016filebench}
V.~Tarasov, E.~Zadok, and S.~Shepler, ``Filebench: A flexible framework for
  file system benchmarking,'' \emph{login: The USENIX Magazine}, vol.~41,
  no.~1, pp. 6--12, 2016.

\bibitem{lsi_raid}
{LSI}, ``{LSI MegaRAID Controller Benchmark Tips},''
  https://docs.broadcom.com/docs/12353177, {Accessed: June 2018}.

\end{thebibliography}
%


\begin{IEEEbiography}[{\includegraphics[width=1in,height=1.304in,clip]{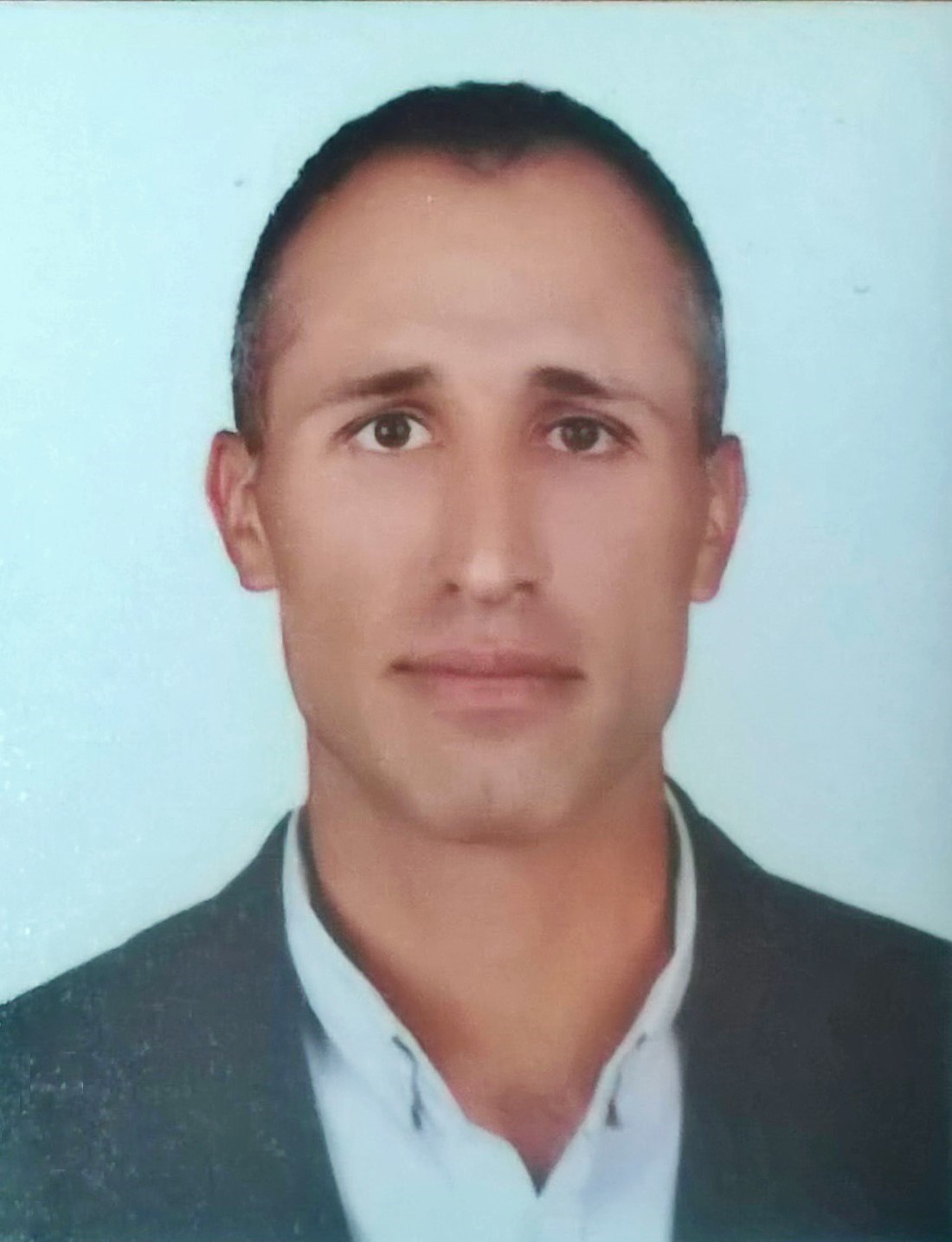}}]
	{Mostafa Kishani}
	received the B.S. degree in computer engineering from Ferdowsi University of Mashhad, Mashhad, Iran, in 2008, M.S. degree in computer Engineering from Amirkabir University of Technology (AUT), Tehran, Iran, in 2010, and PhD degree in computer engineering from Sharif University of Technology (SUT), Tehran, Iran, in 2018. He is currently a postdoctoral fellow in \emph{Data Storage, Networks, and Processing} (DSN) Lab at SUT, Tehran, Iran. 
	He was a hardware engineer in Iranian Space Research Center (ISRC) from 2010 to 2012.
	He was also a member of Institute for Research in Fundamental Sciences (IPM) Memocode team in 2010.
	From September 2015 to April 2016 he was a research assistant in Computer Science and Engineering department of the Chinese University of Hong Kong (CUHK), Hong Kong.
	He was also a research associate in the Hong Kong Polytechnic University (PolyU), Hong Kong, from April 2016 to February 2017.
	
\end{IEEEbiography}


\begin{IEEEbiography}[{\includegraphics[width=1in,height=1.25in,clip,keepaspectratio]{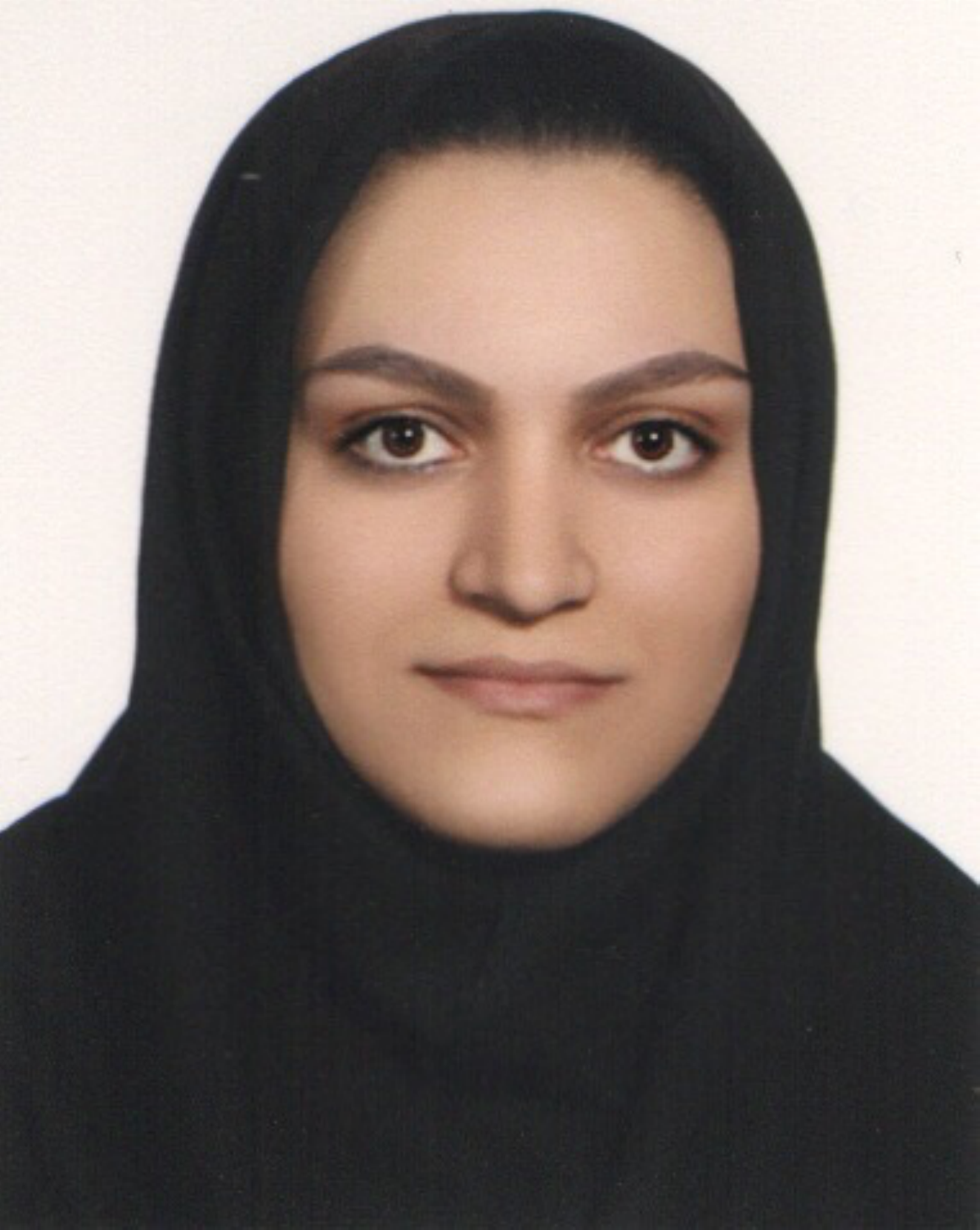}}]{Saba Ahmadian}
	received the B.S. and M.S. degrees in computer engineering {from SUT}, Tehran, Iran, in 2013 and 2015, respectively. From 2011 to 2012, she was a member of \emph{Energy Aware Systems} (EASY) Lab, SUT, where she researched on power reduction techniques on embedded CPUs. {From 2012 to 2015, she was a member of  \emph{Embedded Systems Research} (ESR) Lab, SUT, where she researched on low power and reliability-aware techniques on Automata-based embedded systems.} Currently, she is a Ph.D. candidate at \emph{Data Storage, Networks, and Processing} (DSN) Lab at SUT under supervision of Dr. Hossein Asadi. Her research interests include storage systems design, virtualization platforms, fault tolerant design, and low power systems design.
	
\end{IEEEbiography}


\begin{IEEEbiography}[{\includegraphics[width=1in,height=1.25in,clip]{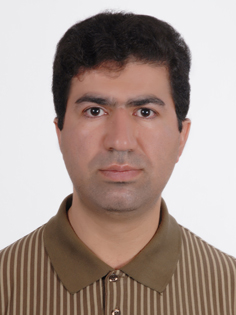}}]
	{Hossein Asadi}
	(M'08, SM'14) received the B.Sc. and M.Sc. degrees in computer engineering from the SUT, Tehran, Iran, in 2000 and 2002, respectively, and the Ph.D. degree in electrical and computer engineering from Northeastern University, Boston, MA, USA, in 2007. 
	
	He was with EMC Corporation, Hopkinton, MA, USA, as a Research Scientist and Senior Hardware Engineer, from 2006 to 2009. From 2002 to 2003, he was a member of the Dependable Systems Laboratory, SUT, where he researched hardware verification techniques. From 2001 to 2002, he was a member of the Sharif Rescue Robots Group. He has been with the Department of Computer Engineering, SUT, since 2009, where he is currently a tenured Associate Professor. He is the Founder and Director of {the DSN Laboratory}, Director of Sharif \emph{High-Performance Computing} (HPC) Center, the Director of Sharif \emph{Information and {Communications} Technology Center} (ICTC), and the President of Sharif ICT Innovation Center. He spent three months in the summer 2015 as a Visiting Professor at the School of Computer and Communication Sciences at the Ecole Poly-technique Federele de Lausanne (EPFL). He is also the co-founder of HPDS corp., designing and fabricating midrange and high-end data storage systems. He has authored and co-authored more than eighty technical papers in reputed journals and conference proceedings. His current research interests include data storage systems and networks, solid-state drives, operating system support for I/O and memory management, and reconfigurable and dependable computing.
	
	Dr. Asadi was a recipient of the Technical Award for the Best Robot Design from the International RoboCup Rescue Competition, organized by AAAI and RoboCup, a recipient of Best Paper Award at the 15th CSI {International} Symposium on \emph{Computer Architecture \& Digital Systems} (CADS), the Distinguished Lecturer Award from SUT in 2010, the Distinguished Researcher Award and the Distinguished Research Institute Award from SUT in 2016, and the Distinguished Technology Award from SUT in 2017. He is also recipient of Extraordinary Ability in Science visa from US Citizenship and Immigration Services in 2008. He has also served as the publication chair of several national and international conferences including CNDS2013, AISP2013, and CSSE2013 during the past four years. Most recently, he has served as a Guest Editor of IEEE Transactions on Computers, an Associate Editor of Microelectronics Reliability, a Program Co-Chair of CADS2015, and the Program Chair of CSI National Computer Conference (CSICC2017). 
\end{IEEEbiography}


\appendix
\begin{appendices}
	\subsection{{Fault injection procedure}}
	\label{sec:app-alg}
	
	{The pseudo-code of the major functions of fault injection implementation is shown in Algorithm~\ref{pseudocode}.  At the beginning of experiment, we initialize the platform variables with the desired parameter values such as block size ($Block\_Chunks$), stripe size ($Array\_Stripes$), number of simulations ($number\_of\_simulations$), and mission time for each iteration ($Simulation\_Time$).} 
	
	{\textbf{Simulation:} In the $Simulation$ procedure, first the SSD pool is generated as detailed in Section IV.C, by calling $SSD\_Pool\_Generator$ procedure. Afterwards, an iteration starts for $number\_of\_simulations$. For each iteration, we simulate an SSD array for the mission time of $Simulation\_Time$. The simulation process starts with initializing the SSD array by choosing \emph{n} random SSDs from SSD pool by calling $Initialize\_SSD\_Array$. Afterwards, $Fault\_Injection$ procedure starts on the SSD array. We finally collect fault injection statistics by calling $Collect\_Statistics$ procedure.} 
	
	{\textbf{Fault\_Injection:} In $Fault\_Injection$ procedure, we perform fault injection on a single SSD array. In a loop, we recognize the next event type (by calling $get\_next\_event\_type$ procedure) and take the appropriate action on reconstruct, scrubbing, and failure events. In the case of failure event, we recognize the device of failure (by calling $Get\_Device\_of\_Failure$ procedure), handle the failure (by calling $Handle\_Failure$ procedure), and generate the next failure for the device of failure (by calling $Generate\_Next\_Failure$), regarding the statistical characteristics of bad chip, bad block, and bad symbol. At the end of each iteration,  the elapsed fault injection time ($Elapsed\_Time$) is set to the next event time (by calling $get\_next\_event\_time$ procedure). Fault injection finishes when $Elapsed\_Time$ reaches $Simulation\_Time$.} 
	
	{The pseudo-code of supplementary functions of fault injection implementation is shown in Algorithm~\ref{pseudocode2}.} 
	
	{\textbf{Handle\_Failure:} This procedure recognizes the failure type and checks the occurrence of ADL, BDL, and SDL throughout the array (by calling $Check\_Stripe\_DL$ procedure).} 
	
	{\textbf{Generate\_Next\_Failure:} This procedure generates the next failure time and type of a specific device, regarding the statistical characteristics of bad chip, bad block, and bad symbol and SSD usage log.} 
	
	{\textbf{Get\_Num\_Faulty\_Chunks:} Returns number of chunks holding at least one affected symbol.}
	
	{\textbf{Get\_Num\_Multi\_Symbol\_Faulty\_Chunks:} Returns number of chunks holding more than one affected symbol (in the case of PMDS, failure of these chunks is handled similar to a device failure). }

	\begin{algorithm*}
		\scriptsize
		\caption{{SSD array fault injection pseudo code.}}\label{pseudocode}
		\begin{algorithmic}
			
			\State {$\#$define $Block\_Chunks = 16$ // number of data chunks per SSD block} 
			\State {$\#$define $num\_of\_simulations = 10,000$ // number of SSD arrays to be simulated} 
			\State {$\#$define $Simulation\_Time = 35,040$ // hours of mission time simulated for each SSD array} 
			\State {$\#$define $Array\_Stripes = 2,097,152$ // number of stripes per SSD array}
			
			\Procedure{Simulation}{}
			\State{// simulates array lifetime for a predefined number of SSD arrays}
			\State {\Call {SSD\_Pool\_Generator(SSD\_Field\_Failure\_Statistics, SSD\_Usage\_Log)}{}}
			\For{$num\_of\_simulations$ }
			\State
			\Call {initialize\_SSD\_Array(SSD\_Pool)}{} // initialize the SSD array
			\State
			\Call {Fault\_Injection(SSD\_Array)}{} // perform fault injection on the SSD array
			\State 
			\Call {Collect\_Statistics()}{} // collect fault injection and failure statistics
			\EndFor
			\EndProcedure
			
			\Procedure{SSD\_Pool\_Generator(SSD\_Field\_Failure\_Statistics, SSD\_Usage\_Log)}{}
			\State{// generates a pool of randomly generated SSDs failure characteristics, regarding the field statistics of SSD failures and SSD usage logs obtained for different workloads}
			\State {$return$ \Call{SSD\_Pool}{}}
			\EndProcedure
			
			\Procedure{initialize\_SSD\_Array(SSD\_Pool)}{}
			\State{// randomly selects array devices from SSD pool}
			\For{$i \in Array\_Devices$}
			\State {$SSD\_Array[i] =$ \Call{SSD\_Pool.Get\_Random\_SSD()}{}}
			\EndFor
			\EndProcedure
			
			\Procedure{Fault\_Injection(SSD\_Array)}{}
			\State{// performs fault injection on a single SSD array}
			\While {$Elapsed\_Time < Simulation\_Time$}
			\If {$get\_next\_event\_type() == reconstruct$} 
			\State
			\Call {Reconstruct()}{} // if the next event is disk reconstruct
			\Else{}{
				\If{$get\_next\_event\_type() == scrubbing$} 
				\State
				\Call {Scrubbing()}{} // if the next event is disk scrubbing
				\Else{}{
					\If{$get\_next\_event\_type() == failure$} // the next event is a failure
					\State{// get target device of the failure}
					\State {$failure\_device\_num =$ \Call{Get\_Device\_of\_Failure()}{}}
					\State{// handle the failure regarding the employed erasure code}
					\State
					\Call {Handle\_Failure(Erasure\_Code)}{} 
					\State{// generate the next failure for the target device}
					\State {$SSDArray[failure\_device\_num].\Call{Generate\_Next\_Failure(Elapsed\_Time)}{}$}
					\EndIf 
				}
				\EndIf 
			}
			\EndIf 
			\State {$Elapsed\_Time = get\_next\_event\_time()$}
			\EndWhile
			\EndProcedure
			
		\end{algorithmic}
	\end{algorithm*}

	\begin{algorithm*}
		\scriptsize
		\caption{SSD {array fault injection pseudo code} (supplementary function definitions)}\label{pseudocode2}
		\begin{algorithmic}
			
			\State {$\#$define $Block\_Chunks = 16$ // number of data chunks per SSD block} 
			\State {$\#$define $Array\_Stripes = 2,097,152$ // number of stripes per SSD array}
			
			\Procedure {Handle\_Failure(Erasure\_Code)}{}
			\State{// handles different types of failure including bad chip, bad block, and bad symbol}
			\If {$\Call{Get\_Next\_Failure()}{}.type == Bad\_Chip$} // in the case of bad chip
			\If {$\Call{Get\_Num\_of\_Failed\_Devices() $>$ 1}{}$}
			\State {$DDF ++$} // increase Double Disk Failure statistics
			\EndIf
			\If {$\Call{Get\_Num\_of\_Failed\_Devices() $>$ 2}{}$}
			\State {$TDF ++$} // increase Triple Disk Failure statistics
			\EndIf
			\State{// check DL incidence for all array stripes}
			\For {$i \in Array\_Stripes$}
			\State {$\Call {Check\_Stripe\_DL(Erasure\_Code, }{}i)$}
			\EndFor
			\EndIf
			\If {$\Call{Get\_Next\_Failure()}{}.type == Bad\_Block$} // in the case of bad block
			\State{// check DL incidence for all stripes the failed block is shared upon}
			\For {$i \in Block\_Chunks$}
			\State {$\Call {Check\_Stripe\_DL(Erasure\_Code, Get\_Failure\_Stripe\_Num()}{}+i)$}
			\EndFor
			\EndIf
			\If {$\Call{Get\_Next\_Failure()}{}.type == Bad\_Symbol$} // in the case of bad symbol
			\State{// check DL incidence in the affected stripe}
			\State {$\Call {Check\_Stripe\_DL(Erasure\_Code, Get\_Failure\_Stripe\_Num()}{})$}
			\EndIf
			\EndProcedure
			
			\Procedure{Generate\_Next\_Failure(Elapsed\_Time)}{}
			\State{// generates the next failure time and type of a specific device, regarding bad chip, bad block, and bad symbol history and SSD characteristics}
			\State{\Call{Set\_Next\_Failure\_Type()}{}} // determine whether the next failure is bad chip, bad block, or bad symbol
			\State{\Call{Set\_Next\_Failure\_Time()}{}} // determine time of the next failure
			\State{\Call{Set\_Next\_Failure\_Location()}{}} // determine location of the next failure
			\EndProcedure
			
			\Procedure{Check\_Stripe\_DL(Erasure\_Code, Stripe\_Number)}{}
			\State{// checks DL incidence in a stripe, regarding the employed erasure code}
			\If {$\Call{Erasure\_Code}{}==RAID5$}
			\State{// in the case of RAID5, if an stripe has more than one faulty chunk, the fault is not correctable}
			\If {$\Call{Get\_Num\_Faulty\_Chunks(Stripe\_Number)}{} > 1$}
			\State {$Stripe\_DL++$} // update statistics of lost stripes
			\EndIf
			\EndIf
			\If {$\Call{Erasure\_Code}{}==RAID6$}
			\State{// in the case of RAID6, if an stripe has more than two faulty chunk, the fault is not correctable}
			\If {$\Call{Get\_Num\_Faulty\_Chunks(Stripe\_Number)}{} > 2$}
			\State {$Stripe\_DL++$} // update statistics of lost stripes
			\EndIf
			\EndIf
			\If {$\Call{Erasure\_Code}{}==PMDS$}
			\State{// in the case of PMDS, if an stripe has more than two faulty chunk, the fault is not correctable}
			\If {$\Call{Get\_Num\_Faulty\_Chunks(Stripe\_Number)}{} > 2$}
			\State {$Stripe\_DL++$} // update statistics of lost stripes
			\Else{}{
				\State{// in the case of PMDS, if the number of faulty chunks is not greater than two, but both faulty chunks contain more than one polluted symbol, the fault is not correctable}
				\If {$\Call{Get\_Num\_Multi\_Symbol\_Faulty\_Chunks(Stipe\_Number)}{} > 1$} 
				\State {$Stripe\_DL++$} // update statistics of lost stripes
				\EndIf
			}
			\EndIf
			\EndIf
			\EndProcedure
			
			\Procedure{Get\_Num\_Faulty\_Chunks(Stripe\_Number)}{}
			\State {// returns number of chunks holding at least one affected symbol}
			\State {$return$ \Call{Num\_Faulty\_Chunks}{}}
			\EndProcedure
			
			\Procedure{Get\_Num\_Multi\_Symbol\_Faulty\_Chunks(Stipe\_Number)}{}
			\State {// returns number of chunks holding more than one affected symbol}
			\State {$return$ \Call{Num\_Multi\_Symbol\_Faulty\_Chunks}{}}
			\EndProcedure
		\end{algorithmic}
	\end{algorithm*}

	\subsection{Performance Comparison}
	\label{sec:perf}
	In this {section,} we evaluate the performance of different types of erasure code. To this end, we run the workloads of FIO and Filebench on the SSDs array (the detailed information about workloads parameters is provided in Section \ref{sec:experimental setup}). 
	For FIO workloads with \emph{random} access pattern, the performance is reported in terms of \emph{I/O Operations per Second} (IOPS) and latency. For FIO workloads with \emph{sequential} access pattern, however, we report \emph{Bandwidth} (BW). 
	Finally, for Filebench workloads we report IOPS, latency, and BW. 
	
	\begin{figure}[!h]
		\centering
		\subfloat[IOPS (for the workloads with random access pattern)]{\includegraphics[width=0.45\textwidth]{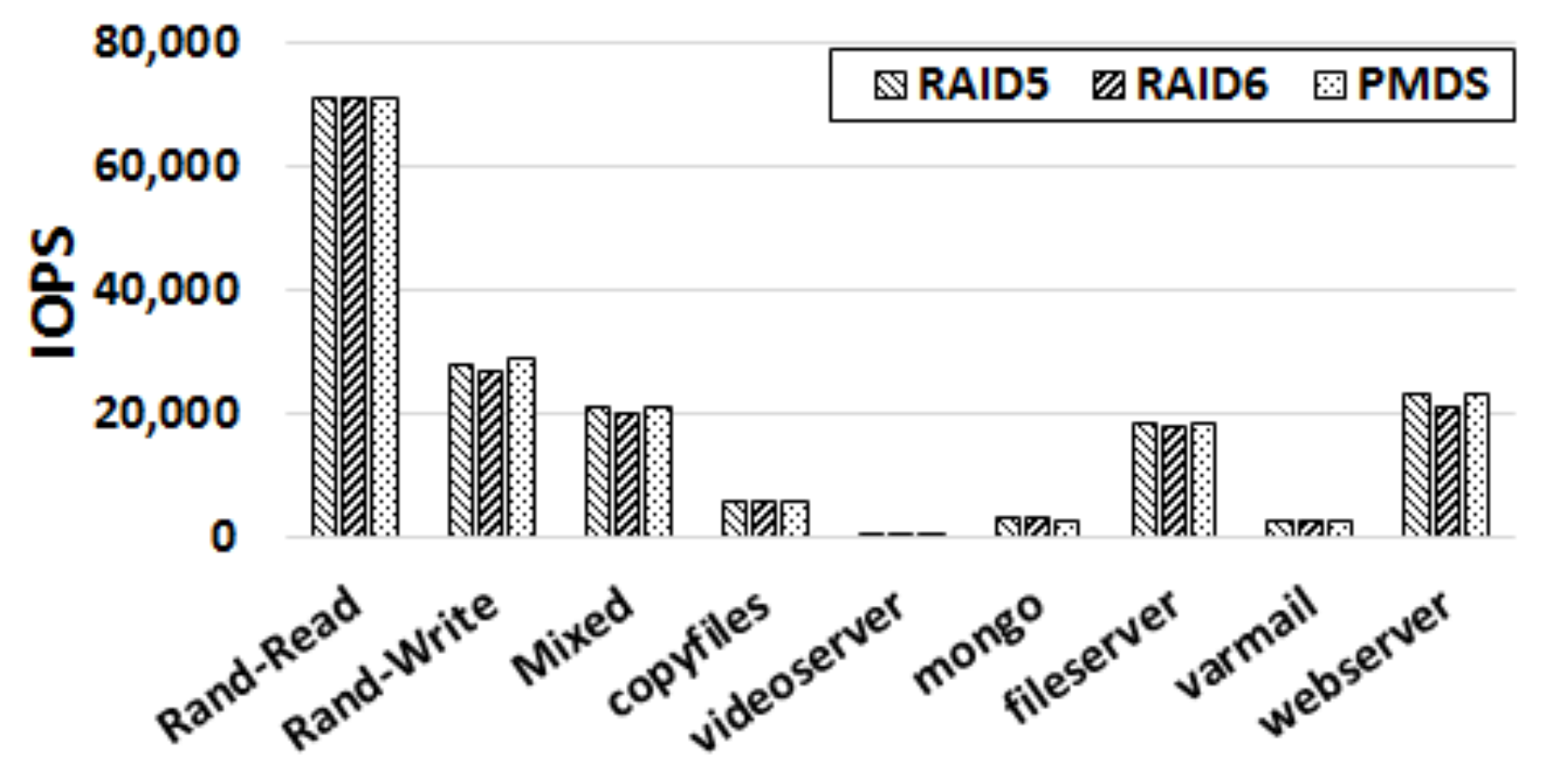}%
			\label{fig:iops}}
		\hfil
		\hspace{-.8pt}
		\subfloat[Latency (for the workloads with random access pattern)]{\includegraphics[width=0.45\textwidth]{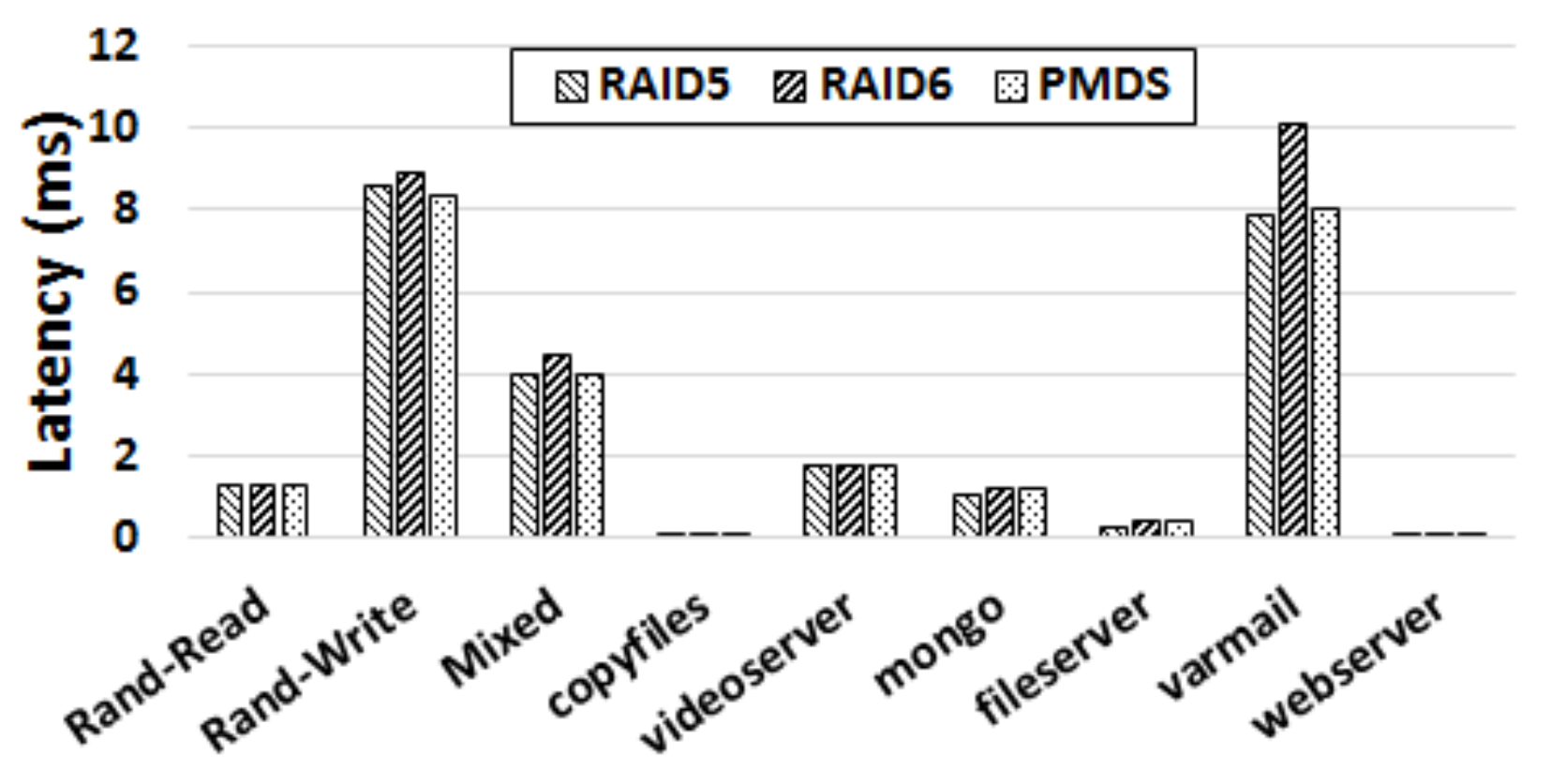}%
			\label{fig:latency}}
		\hfil
		\hspace{-.8pt}
		\subfloat[BW (for the workloads with sequential access pattern)]{\includegraphics[width=0.45\textwidth]{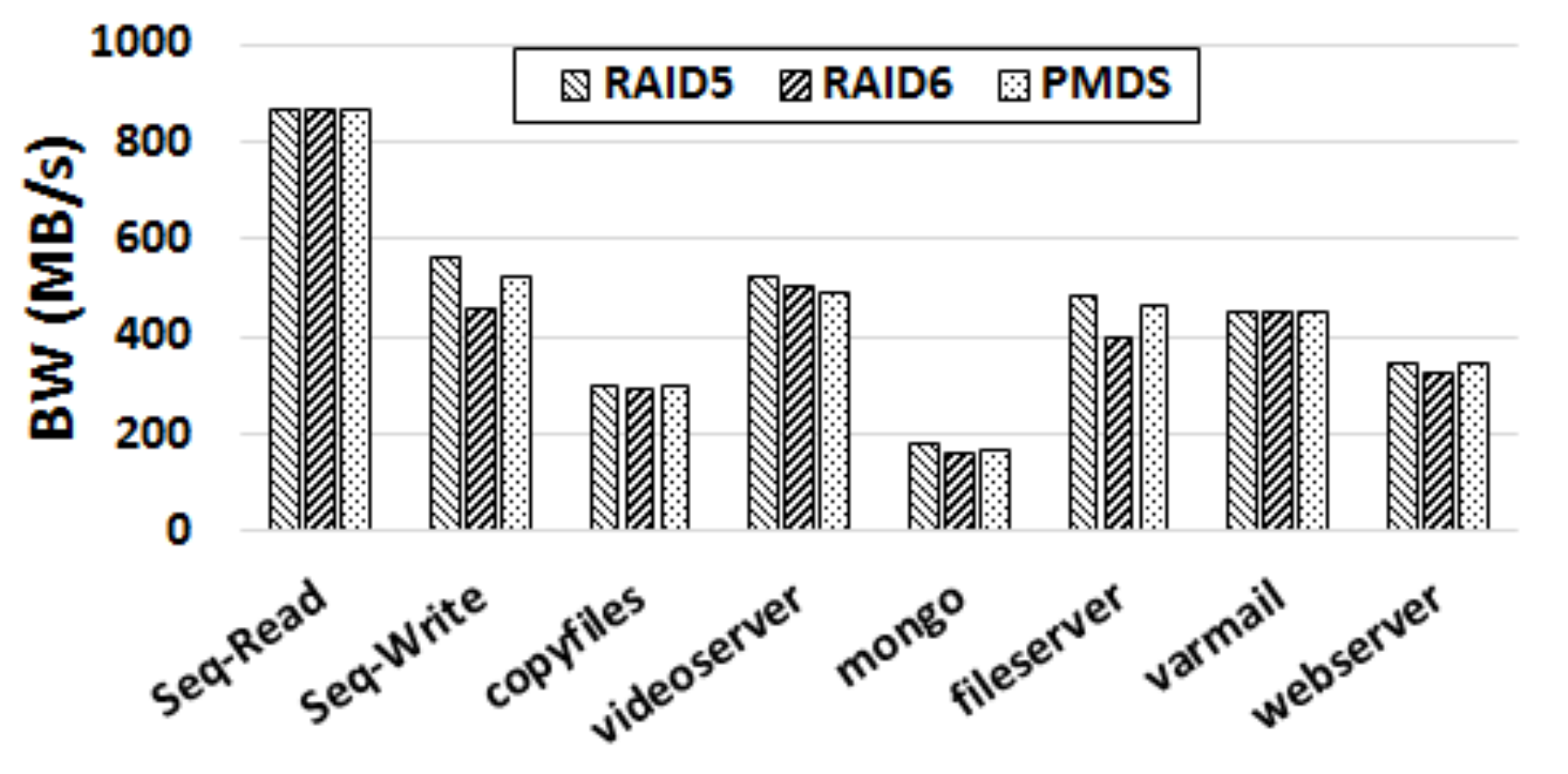}%
			\label{fig:bw}}
		
		\caption{Comparing {performance} for different workloads.}
		
		\label{fig:perf-result}
	\end{figure}
	
	Fig. \ref{fig:perf-result} shows the performance results of the experiments for RAID5, RAID6, and PMDS configuration  (Fig.~\ref{fig:iops}, Fig.~\ref{fig:latency}, and Fig.~\ref{fig:bw} report IOPS, latency, and BW, {respectively}).
	Regarding the empirical results, we have five main observations:
	\begin{itemize}[leftmargin=*]
		\item
		The maximum performance is achieved in the workloads with read requests including Rand-Read, Seq-Read, and Video server, due {to different service times} for read and write requests.
		\item
		In Rand-Read, Seq-Read, and Video server  workloads{, we} get the similar performance from RAID5, RAID6, and PMDS configurations. 
		This observation is described by the fact that the mentioned workloads only contain read requests, mandating no parity calculations. {Hence,} all examined erasure codes employing the same number of physical SSDs can read the data with the same bandwidth. 
		\item
		In the majority of workloads, RAID5 and PMDS achieve {higher} performance than RAID6.
		We observe some exceptions such as \emph{Video server} BW in which RAID6 performs slightly better than PMDS.
		\item
		In the workload dominated by write operations such as Rand-Write, Seq-Write, and Varmail, we observe that both RAID5 and PMDS achieve a considerably higher performance than RAID6, due to the significant parity overhead of RAID6 configuration. 
		\item
		In the Mixed and Fileserver workloads, both characterized by the mixture of random read and write requests, we observe a greater latency compared to fully random read (Rand-Read) and lower latency compared to fully random write (Rand-Write) workloads. In this type of workload (i.e., the mixture of random read and {write),} the performance of RAID5 is {higher} than both RAID6 and PMDS.
	\end{itemize}
	
	\subsection{Endurance}
	\label{sec:endurance}
\begin{figure}
		\centering
		
		\includegraphics[width=0.5\textwidth]{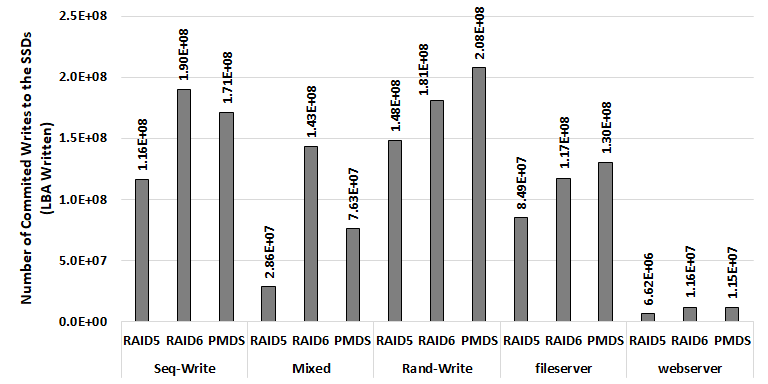}
		\caption{Comparing the impact of different erasure codes on SSD {endurance.}}
		\label{fig:endurance}
		
	\end{figure}

	Fig.~\ref{fig:endurance} compares the impact of different erasure codes on SSD endurance. 
	As a representative for SSD endurance, number of LBAs written to the array {is} collected for each individual SSD using S.M.A.R.T tool, while Fig.~\ref{fig:endurance} reports the average of LBAs written to SSDs. 
	The impact of erasure codes on endurance is considerably correlated with workload. 
	While RAID5 always performs better than both RAID6 and PMDS in terms of endurance (when deploying RAID5, RAID6, and PMDS arrays with equal raw capacity, as we do), relative endurance of PMDS and RAID6 is subject to workload.
	As analyzed in Section~\ref{sec:background} (Table~\ref{tab:update penalty comparison}), PMDS imposes greater overhead than RAID6 in single sector and single row updates, while in the case of full stripe updates, it performs better than RAID6.
	PMDS codes, however, always perform worse than RAID5, as shown in Fig.~\ref{fig:endurance} and also predicted in our analysis in Section~\ref{sec:background}, due to an extra global parity overhead.

\end{appendices}

\end{document}